\documentclass[final,5p,times,twocolumn,sort&compress]{elsarticle}

\usepackage{etoolbox}

\usepackage{amsmath}
\usepackage{amsfonts}
\usepackage{amssymb}
\usepackage[usenames,dvipsnames]{color}
\usepackage[colorlinks=true,linkcolor=Maroon,citecolor=Maroon,urlcolor=Maroon]{hyperref}
\usepackage{graphicx}
\usepackage{subcaption}
\usepackage{xcolor}
\usepackage{comment}
\usepackage{multirow}
\usepackage{booktabs}

\usepackage{cuted}

\makeatletter
\def\ps@pprintTitle{%
 \let\@oddhead\@empty 
 \let\@evenhead\@empty
 \def\@oddfoot{}%
 \let\@evenfoot\@oddfoot}
 \makeatother 

\newcommand{\n}{\mathbf{n}}
\newcommand{\x}{\mathbf{x}}

\newcommand{\f}{\mathbf{f}}
\newcommand{\g}{\mathbf{g}}
\newcommand{\vv}{\mathbf{v}}
\newcommand{\uu}{\mathbf{u}}
\newcommand{\z}{\mathbf{z}}
\newcommand{\A}{\mathbf{A}}
\newcommand{\D}{\mathbf{D}}

\newcommand{\J}{\mathbf{J}}
\newcommand{\I}{\mathbf{I}}
\newcommand{\M}{\mathbf{M}}
\newcommand{\N}{\mathbf{N}}
\newcommand{\PP}{\mathbf{P}}
\newcommand{\OO}{\mathbf{0}}

\newcommand{\U}{\mathbf{U}}
\newcommand{\V}{\mathbf{V}}

\newcommand{\nn}{\boldsymbol{\hat{n}}}
\newcommand{\rr}{\boldsymbol{\hat{r}}}
\newcommand{\pp}{\boldsymbol{\hat{p}}}
\newcommand{\et}{\boldsymbol{\eta}}

\newcommand{\gsim}{\overset{\sim}{\g}}
\newcommand{\Gam}{\boldsymbol{\Gamma}}
\newcommand{\Lamb}{\boldsymbol{\Lambda}}
\newcommand{\Sig}{\boldsymbol{\Sigma}}
\newcommand{\sig}{\boldsymbol{\sigma}}

\newcommand{\mw}[1]{\langle #1 \rangle}

\newcommand{\SM}{(\textit{SM})}

\begin{document}

    \begin{frontmatter}

        \title{
            A Unifying Framework for Amplification Mechanisms:\\
            Spectral Criticality, Resonance and Non-Normality
        }

        \author[add1]{Virgile Troude}
        \author[add1]{Didier Sornette}

        \address[add1]{\scriptsize
            Institute of Risk Analysis, Prediction and Management (Risks-X),\\
            Academy for Advanced Interdisciplinary Sciences,\\
            Southern University of Science and Technology, Shenzhen, China
        }

        \begin{abstract}
        We bring together three key amplification mechanisms in linear dynamical systems: spectral criticality, resonance, and non-normality. 
        We present a unified linear framework that both distinguishes and quantitatively links these effects through two fundamental parameters: (i) the spectral distance to a conventional bifurcation or to a resonance and (ii) a non-normal index $K$ (or condition number $\kappa$) that measures the obliqueness of the eigenvectors. Closed-form expressions for the system's response in the form of the variance $v_\infty$ of the observable responding to both Gaussian noise and periodic forcing reveal a general amplification law  $v_\infty = v_0 \left( 1 + \mathcal{G}(K) \right)$ with non-normal gain
$\mathcal{G}(K) \propto K^2$  represented in universal phase diagrams. 
        By reanalyzing a model of remote earthquake triggering based on breaking of Hamiltonian symmetry, we illustrate how our two-parameter framework significantly expands both the range of conditions under which amplification can occur and the magnitude of the resulting response, revealing a broad pseudo-critical regime associated with large $\kappa$ that previous single-parameter approaches overlooked.
        Similarly, in the Non-Hermitian extensions of quantum optics provided by Forward Four-Wave Mixing (FFWM) experiments,
        we show the presence of a counterintuitive gain-from-loss effect that directly manifests non-normal amplification in a propagating-wave setting. 
        This predicts the possibility to engineer transient optical energy amplification without the need for true lasing or exact $\mathcal{PT}$-symmetry breaking.
       Our framework applies to many other physical, natural and social systems and offers new diagnostic tools to distinguish true critical behavior from transient amplification driven by non-normality. 
\end{abstract}

    \end{frontmatter}


    Many natural and engineered systems exhibit sudden, disproportionate responses to small disturbances or subtle shifts in control parameters—well before classical indicators like spectral instability or resonance emerge. Earthquake ruptures, hydrodynamic bursts, and financial crashes are striking examples where minor perturbations can trigger massive departures from equilibrium. Such abrupt, switch-like transitions are pervasive across complex systems: epileptic seizures in the brain \cite{Maturana2020,Royer2022}, ecosystem collapses \cite{Tang2014,Hirota2011}, market crashes \cite{Sornette2017}, and rapid epigenetic reprogramming \cite{busto2020stochastic}, all involve transient amplifications that vastly exceed background dynamics. These phenomena highlight the need for a broader framework to understand how small inputs can drive large-scale responses.

    Classical bifurcation theory seeks to explain such events by the slow drift of a control parameter until an 
    eigenvalue of the linearised Jacobian $\J$ crosses the imaginary axis,
    generating well-known early-warning signals such as critical slowing-down and rising variance \cite{Scheffer2009}.  
    Two enduring puzzles, however, limit this picture:
    (\emph{i}) critical points are measure-zero targets in all dimensional parameter spaces,
    and (\emph{ii}) critical-like signatures are frequently observed far from any spectral instability \cite{troude2024}.

    When $\J$ is non-normal ($[\J,\J^\dag]\neq0$), the $\varepsilon$-pseudospectrum 
    $\sigma_\varepsilon(\J)=\{\lambda\in\mathbb{C}\;|\;\|(\J-\lambda I)^{-1}\|\ge\varepsilon^{-1}\}$
    can inflate dramatically \cite{Embree2005}.  
    The eigenbasis condition number $\kappa$ (defines as the largest singular value divided by the smallest singular value of the eigenbasis transformation matrix) quantifies this inflation;
    for $\kappa\gg1$ a finite perturbation is transiently amplified proportionally to $\kappa^{2}$ \cite{troude2024}.  
    These pseudo-critical transients mimic all the hallmarks of true criticality, even though the real parts of the eigenvalues remain negative. In essence, non-normality unfolds a continuum of critical-like states surrounding the bifurcation manifold.
    Pseudo-critical bursts have been reported in balanced neural networks \cite{Murphy2009}, hierarchical food-webs \cite{OBrien2021}, turbulent shear flows \cite{Trefethen1993}
    and in minute-scale DNA-methylation surges \cite{troude2025}.  
    The ubiquity of non-normal amplification calls for a comprehensive theoretical approach.

    Here, we develop such a unified analytical framework that integrates three distinct amplification mechanisms—(i) critical bifurcations, (ii) resonant amplification in underdamped systems, and (iii) pseudo-critical transients driven by non-normality—into a single coherent formalism with compatible concepts and notations.
    By combining pseudospectral analysis with condition-number geometry, we
    (a) derive closed-form criteria that clearly distinguish these regimes;
    (b) quantify their joint effects on system responses; and
    (c) offer practical diagnostic tools—such as pseudospectral growth bounds and in-situ estimates of $\kappa$—to detect pseudo-critical amplification in both laboratory and field settings.

    This framework charts the full landscape of dynamical amplification, revealing how non-normality opens a continuum of critical-like states even when eigenvalues remain stable, and providing a comprehensive guide for interpreting large fluctuations across disciplines.

    \section*{Generalised Dynamical System}
    
    The following general setting captures these three amplification routes.  
    Consider an $n$-dimensional state vector $\x(t)\in\mathbb{R}^{n}$ governed by the generalised Langevin equation
    \begin{equation}
    \ddot{\x}+\gamma\dot{\x}= \f(\x)+\g(t),
    \qquad 
    \gamma>0 ,
    \label{eq:GLE}
    \end{equation}
    where $\gamma$ is an isotropic damping coefficient and $\g(t)$ is an external forcing to be specified
    e.g. Gaussian white noise or sinusoidal forcing. In full generality. 
    the force field $\f(\x)$ can be non-variational.
    Employing the Helmholtz decomposition\,\cite{glotz2023}, it can expressed as
    \begin{equation}
    \f(\x)= -\nabla U(\x) \;+\;
            \bigl(\nabla^{\!\dag} \A(\x)\bigr)^{\!\dag},
    \qquad 
    \A^\dag(\x)=-\A(\x),
    \label{eq:Helmholtz}
    \end{equation}
    where $U(\x)$ captures the conservative contributions while the solenoidal term arising from the anti-Hermitian
    tensor $\A(\x)$ generates circulation in phase space and breaks detailed balance. 
    
    Expanding \eqref{eq:GLE} around a stable equilibrium $\x^\star\!=\!0$ yields the linear system
    \begin{equation}
        \ddot{\x}+\gamma\dot{\x}=\J\x+\g(t),
        \qquad 
        \J:=\bigl[\nabla\f(\x)\bigr]_{\x=0},
        \label{eq:lin}
    \end{equation}
    where stability implies that all eigenvalues of $\J$ have negative real part.
    The component $\bigl(\nabla^{\!\dag} \A(\x)\bigr)^{\!\dag}$ of the force \eqref{eq:Helmholtz} is responsible
    for non-normality of $\J$, which appears naturally as a large sub-class of the non-variational family.
    Non-normality implies that $\J$ cannot be diagonalized by a unitary transformation.
   
    The intuition behind this model is to view it as a system of coupled, damped oscillators, where each oscillator interacts in a non-symmetric and hierarchical manner. We consider stable and non-normal systems,
    meaning $\J$ is non-normal and has eigenvalues with negative real part.
    The degree of non-normality is quantified by the condition number $\kappa$ of the eigenbasis transformation matrix $\PP$ defined through the eigen decomposition of $\J$ as
    $\J = \PP \Lamb \PP^{-1}$, where $\Lamb = \text{Diag}(\lambda \mid \lambda \in \sig(\J))$,
    with $\sig(\J)$ the spectrum of $\J$.
    The condition number $\kappa$ is defined as the ratio between the largest and smallest singular values of $\PP$.
    A condition number $\kappa = 1$ means that $\J$ is normal,
    while $\kappa > 1$ characterizes a non-normal system.
    Strongly non-normal systems ($\kappa \gg 1$) can exhibit dynamical behaviors that closely resemble those near true bifurcations, despite being far from any actual bifurcation point, a phenomenon that has been referred to as a ``pseudo-bifurcation'' \cite{troude2024}.

    A simplified scenario that nevertheless captures the essence of non-normality is provided by a matrix $\PP$ 
    whose singular values are all equal to $1$, except for the smallest one, which is $1/\kappa$.
    The smallness of this singular value directly determines the magnitude of the condition number $\kappa$, 
    which quantifies the degree of non-normality in the system. In this simplified case, 
    the matrix $\PP$ is approximately unitary in all directions except along one dimension characterized by a specific non-normal mode $\nn$.
    It has been shown \cite{troude2024} that, up to a unitary transformation, one can write
    $\PP = \I + (\kappa^{-1}-1)\nn\nn^\dag$, and correspondingly,
    $\PP^{-1} = \I + (\kappa - 1)\nn\nn^\dag$, with $\I$ being the identity matrix.
    This specific structure reveals that excitations along the non-normal mode $\nn$ induced by external forcing will experience amplification by a factor $\kappa$,
    a hallmark of pseudo-critical behavior.

    For later reference, we assume the system is damped, so that the homogeneous solution decays to zero and only the particular solution remains (see Supplementary Material, \SM). The resulting solution is given by
    \begin{equation}
    \x(t)=\int_{0}^{t}\! G(\J,t-s)\,\g(s)\,ds
    .
    \label{eq:gen_solution}
    \end{equation}
    
    When system \eqref{eq:GLE} is driven by an additive Gaussian white noise
    \begin{equation}
        \g(t) = \sqrt{2\delta}\,\boldsymbol{\eta}(t),
        \qquad
        \langle\eta_{i}(t)\,\eta_{j}(s)\rangle = \delta_{ij}\,\delta(t-s),
        \label{eq:whitenoise}
    \end{equation}
    with noise variance $\delta$,
    since the deterministic contribution from the initial condition decays exponentially,
    the stationary statistics are governed by the particular solution $\x$ \eqref{eq:gen_solution}.
    The long-term mean-squared deviation (MSD) is then given by \SM
    \begin{equation}
        v_\infty := \bigl\langle\|\x(t)\|^{2}\bigr\rangle
        = 2\delta \!\int_{0}^{\infty}\!\!\|\,G(\J,\tau)\|_{F}^{2}\,d\tau.
        \label{eq:vinf_def}
    \end{equation}
    where $\|G(\J,t)\|_F$ is the Froebenius norm of $G(\J, t)$.
    
        Using the spectral decomposition of $\J$ and
    \(
    \PP = \I + (\kappa^{-1} - 1)\,\nn\nn^\dag, \, \nn \sim (1,\,1)/\sqrt{2},
    \)
    in the two-dimensional case with a unique non-normal mode,
    $\|G(\J,\tau)\|_{F}^{2}$ can be written as~\SM
    \begin{equation}
        \|G(\J, t)\|_F^2 =
        \left|G(\lambda_+,t)\right|^2
        + \left|G(\lambda_-,t)\right|^2
        + K^2 \left|G(\lambda_+,t) - G(\lambda_-,t)\right|^2,
        \label{eq:2d-frob}
    \end{equation}
    where the degree of non-normality is defined as
    \begin{equation}
        K := \frac{1}{2}\bigl|\kappa - \kappa^{-1}\bigr|.
        \label{eq:Kindex}
    \end{equation}
    The third term in \eqref{eq:2d-frob}, which vanishes for $K=0$ ($\kappa = 1$),
    encodes transient energy bursts that enhance the MSD by a factor proportional to~$K^2$.

    Let us define  $\theta(\lambda_i)=\sqrt{\bigl(\tfrac{\gamma}{2}\bigr)^{2}+\lambda_i}$, with $\lambda_i\in\sig(\J)$.
    Starting from the spectral decomposition $\J = \PP \Lamb \PP^{-1}$ of $\J$,
    then $G(\J, t) = \PP G(\Lamb, t) \PP^{-1}$ is the spectral decomposition of $G(\J, t)$, 
    where $G(\Lamb, t)$ is the diagonal matrix with elements
    \begin{equation}
        G(\lambda_i,t)=e^{-\gamma t/2}\,   
        \frac{\sinh\!\bigl(\theta(\lambda_i)t\bigr)}{\theta(\lambda_i)}~~, ~~~\lambda_i\in\sig(\J)~.
    \label{eq:G_G0} 
    \end{equation}
    When $\J$ is non-normal ($[\J,\J^\dag]\neq0$), the kernels exhibit
    direction-dependent transient growth with amplitude quantified by a degree of non-normality $K$ (\ref{eq:Kindex}),
    dependent on the condition number $\kappa$ of the eigenbasis transformation.
     For a normal matrix ($\kappa = 1$), the degree of non-normality is $K=0$.
    The Froebenius norm $\|G(\J,t)\|_F$ of $G(\J, t)$  (where $\|\J\|_F = \sqrt{\text{Tr}(\J^\dag\J)}$ is the Frobenius norm of matrix $\J$),
    may therefore increase as the result of a competition between exponentials that can create a finite-time maximum 
    before ultimately decaying, a hallmark of non-normal amplification.

    \section*{Reduced Non-Normal Forms}
    
    Our aim in this section is to simplify the parametrization of a two-dimensional system 
    with a real Jacobian matrix so that its stability, degeneracies, and non-normality can be analyzed in the most transparent way.

A real $2\times 2$ Jacobian has eigenvalues that are either both real or form a complex-conjugate pair. We write them in the unified form
\begin{equation}
        \lambda_\pm = \lambda \pm \Delta \chi ,
\end{equation}
    where $\lambda, \Delta \in \mathbb{R}$ and $\chi \in \{ 1, i \}$ distinguishes the purely real ($\chi = 1$) from the complex-conjugate ($\chi = i$) case.
   In both cases, the spectrum depends on three control parameters $(\lambda, \Delta, \kappa)$, with $\kappa$ measuring the degree of non-normality
   
   However, treating all three parameters simultaneously can obscure the essential control mechanisms.
To disentangle them, we introduce two alternative time rescalings, each leaving only one independent parameter in the spectrum:
   \begin{subequations} \label{eq:time_form}
        \begin{align}
            \alpha &= -\frac{\lambda}{\Delta}, & t &\to t\,\Delta, & &\text{(\textbf{$\alpha$-form})}, \\
            \beta  &= -\frac{\Delta}{\lambda}, & t &\to t\,|\lambda|, & &\text{(\textbf{$\beta$-form})}.
        \end{align}
    \end{subequations}
    We assume stability, $\lambda < 0$, and, without loss of generality, $\Delta > 0$. The corresponding eigenvalues are then
  \begin{equation} \label{eq:eigs_alpha_beta}
        \lambda_\pm =
        \begin{cases}
            -\alpha \pm \chi, & \text{(\textbf{$\alpha$-form})}, \\
            -1 \pm \beta\chi, & \text{(\textbf{$\beta$-form})}.
        \end{cases}
    \end{equation}
    Up to a unitary transformation and the above time rescaling in \eqref{eq:time_form},  
    the system's Jacobian $\J$ in \eqref{eq:lin} reduces to (see SM)
    \begin{equation}
        \Gamma^\alpha_{\chi} =
        \begin{pmatrix}
            -\alpha & \chi \kappa \\
            \chi \kappa^{-1} & -\alpha
        \end{pmatrix},
        \quad
        \Gamma^\beta_{\chi} =
        \begin{pmatrix}
            -1 & \chi \beta \kappa \\
            \chi \beta \kappa^{-1} & -1
        \end{pmatrix}.
    \end{equation}
    
    The $\alpha$-form normalizes time to the spectral gap $\Delta$ between the two eigenvalues.
    The  dynamics is measured in units of the inverse splitting between eigenvalues. This makes $\alpha$ a distance-to-criticality parameter,
    which is best suited for tracking how non-normality interacts with critical slowing down near a bifurcation.
    The $\beta$-form normalizes time to the mean decay rate $|\lambda|$ of the modes.
The  dynamics is measured in units of the inverse average damping time. This makes $\beta$ a degeneracy parameter,
best suited for describing smooth spectral transitions and mode collisions, rather than approaching an instability threshold.
   
   \vskip 0.3cm
  $\bullet$  {\bf Overdamped dynamics}:
  \begin{equation}
        \dot{\mathbf{x}} = \Gamma^{\alpha/\beta}_{\chi} \, \mathbf{x}~.
        \label{hwtbqbuk}
    \end{equation}
In this case, criticality occurs when $\Re(\lambda_\pm) \to 0^-$.
In the $\alpha$-form, this corresponds to $\alpha \to \alpha_c^+$, with
$\alpha_c=1$ for $\chi=1$ and $\alpha_c=0$ for $\chi=i$.
Thus, $\alpha$ directly measures proximity to criticality, while $\kappa$ modulates the impact of non-normality.
Changing $\chi$ switches between oscillatory and purely exponential damping without altering stability.
    
      \vskip 0.3cm
  $\bullet$  {\bf Underdamped dynamics}:
 \begin{equation}
        \ddot{\mathbf{x}} = \Gamma^{\alpha/\beta}_{\chi} \, \mathbf{x}~.
 \end{equation}
  The four eigenvalues are $\pm \sqrt{\lambda_\pm}$.  
    In the $\alpha$-form, this gives
    \(
        \pm i \sqrt{\alpha \pm \chi}.
    \)
    For $\chi = 1$, a bifurcation occurs at $\alpha \to 1^+$,  similar to the overdamped case.  
    For $\chi = i$, the system is always unstable: switching from $\chi = 1$ to $\chi = i$ then induces a discontinuous bifurcation.  
    Thus, the $\alpha$-form is less suited to describing smooth transitions from real to complex-conjugate eigenvalues.
  For such smooth transitions, the $\beta$-form is more appropriate:  
    eigenvalues collide when $\beta = 0$, corresponding to a degeneracy point of the system.
 
    The two reduced forms serve complementary purposes:
    \begin{itemize}
        \item The \textbf{$\alpha$-form} parametrizes the distance from criticality.
        \item The \textbf{$\beta$-form} captures degeneracy and smooth transitions between real and complex-conjugate eigenvalues.
    \end{itemize}
    Together, they form a minimal yet complete framework for analyzing how non-normality combines with criticality or resonance to produce amplification phenomena.
    
    \section*{Overdamped System Driven by a Gaussian White Noise}

We begin with this simple case for pedagogical clarity, as it provides intuition for the more general scenarios discussed later.
For this, we quantify how non-normality and criticality influence the response to external forcing when the system \eqref{eq:GLE}, 
reduced to (\ref{hwtbqbuk}) for its deterministic part, is driven by additive Gaussian white noise \eqref{eq:whitenoise}.
The long-term mean-squared deviation (MSD) then takes the form given in equation (\ref{eq:vinf_def}).
  
    The overdamped limit of~\eqref{eq:GLE}, i.e.
    \(
    \gamma\dot{\x} = \f(\x) + \g(t),
    \)
    is recovered for $\gamma \gg \lambda_\pm$.
    In this regime, the propagator simplifies to $G(\J,t) \to e^{\J t/\gamma}$,
    maintaining a direct link to analytically tractable overdamped models
    studied in \cite{troude2024,troude2025}.

    Transient growth in the response
    -- defined by a non-monotonically decaying kernel $\|G(\J, t)\|_F^2$ --
    occurs for $K > K_c \ge 0$.
    For the $\alpha$- and $\beta$-form, thresholds are given by:
    \begin{subequations} \label{eq:z_c}
        \begin{align}
            &\text{(\textbf{$\alpha$-form})} 
            &&K_c = 
            \begin{cases}
                \sqrt{\dfrac{\sqrt{\alpha^2 - 1}}{\alpha - \sqrt{\alpha^2-1}}},\;\alpha > 1, &\chi = 1, \\[6pt]
                \sqrt{\dfrac{\alpha}{\sqrt{\alpha^2+1} - \alpha}},\;\alpha > 0, &\chi = i,
            \end{cases} \\[4pt]
            &\text{(\textbf{$\beta$-form})} 
            &&K_c = 
            \begin{cases}
                \sqrt{\dfrac{\sqrt{1 -\beta^2}}{1 - \sqrt{1-\beta^2}}},\; 1 > \beta \ge 0, &\chi = 1, \\[6pt]
                \sqrt{\dfrac{1}{\sqrt{\beta^2+1} - 1}},\;\beta > 0, &\chi = i.
            \end{cases}
        \end{align}
    \end{subequations}
    
    \begin{itemize}
        \item For $K > K_c$ (with $K_c$ depending on $\alpha$ or $\beta$ and $\chi$),  
        the system exhibits amplified fluctuations, qualifying as \emph{pseudo-critical},  
        since it temporarily mimics the behavior of a truly critical system.
        \item For $K < K_c$, no amplification occurs,
        and Gaussian perturbations decay exponentially as in a normal system, 
        indicating resilience to external forcing.
    \end{itemize}
    We can summarize the main observations as:
    \begin{itemize}
        \item \textbf{\(\alpha\)-form:}  
        Near criticality, $K_c \to 0$, the system displays pseudo-critical behavior before becoming truly critical.  
        Far from criticality, a useful heuristic is $K > \sqrt{2}\,\alpha$ ($\alpha \gg 1$) for transient amplification to occur.

        \item \textbf{\(\beta\)-form:}  
        As $\beta \to 0$ (degeneracy), $K_c \approx \sqrt{2}/\beta \to \infty$;
        here $\J$ approaches a scalar multiple of the identity,
        making it invariant under basis changes and suppressing non-normal effects.  
        Away from degeneracy, $K_c \to 0$, and non-normal amplification is generally present.
    \end{itemize}

    In summary, within the Gaussian overdamped framework,
the closer the system is to criticality, the easier it is to observe pseudo-critical transient amplification.
Moreover, near eigenvalue degeneracy, the system becomes insensitive to non-normal effects.
    Figure \ref{fig:phase_gen} (left column) illustrates these pseudo-critical regions in $(K,\alpha)$ and $(K,\beta)$ diagrams,  
    which we refer to as \emph{phase diagrams} since the structural response changes qualitatively as $K$ crosses $K_c$,  
    depending on the approach to either criticality or degeneracy.

    \section*{General System Driven by a Gaussian White Noise}

    We now quantify how non-normality and criticality influence the long-time variance when the
    system \eqref{eq:GLE} is driven by an external additive Gaussian white noise \eqref{eq:whitenoise}.
            
    In the two-dimensional case, where $\J$ is a real matrix, the eigenvalues $\lambda_\pm$ 
    for the $\alpha$- and $\beta$-forms are given
    by~\eqref{eq:eigs_alpha_beta}. The kernel is expressed in~\eqref{eq:G_G0},
    and the variance can be computed from~\eqref{eq:vinf_def} and~\eqref{eq:2d-frob}.
    From these relations, we obtain the following expressions for the MSD \SM
    \begin{subequations}    \label{eq:var_form}
        \footnotesize
        \begin{align}
            &\text{(\textbf{$\alpha$-form})}
            &&v_\infty = 
            \begin{cases}
                \dfrac{2\delta}{\gamma} \dfrac{\alpha}{\alpha^2 - 1}
                \left[ 1 + K^2 \dfrac{\gamma^2 + \alpha}{\alpha(\gamma^2 \alpha + 1)} \right],
                \, \alpha > 1, &\chi = 1, \\
                \dfrac{2\delta}{\gamma} \dfrac{1}{\alpha - \gamma^{-2}}
                \left[ 1 + K^2 \dfrac{\gamma^2 + \alpha}{\gamma^2 (\alpha^2 + 1)} \right],
                \,\alpha > \gamma^{-2}, &\chi = i,
            \end{cases} \\
            &\text{(\textbf{$\beta$-form})}
            &&v_\infty = 
            \begin{cases}
                \dfrac{2\delta}{\gamma} \dfrac{1}{1 - \beta^2}
                \left[ 1 + K^2 \beta^2 \dfrac{1 + \gamma^2}{\gamma^2 + \beta^2} \right],
                \, 1 > \beta > 0, &\chi = 1, \\
                \dfrac{2\delta}{\gamma} \dfrac{1}{1 - (\beta/\gamma)^2}
                \left[ 1 + K^2 \left( \frac{\beta}{\gamma} \right)^2
                \dfrac{1 + \gamma^2}{1 + \beta^2} \right],
                \, \gamma > \beta > 0, &\chi = i.
            \end{cases}
        \end{align}
    \end{subequations}
    In all cases, the MSD exhibits the general form
    \begin{equation}    \label{eq:uni_form_var}
        v_\infty = v_0 \left( 1 + \mathcal{G}(K) \right),
        \quad \mathcal{G}(K) = K^2\, \Delta v,
    \end{equation}
    where $v_0$ is the MSD of the corresponding \emph{normal} system (i.e., $K = 0$).
    We refer to $\mathcal{G}(K)$ as the \emph{non-normal gain},
    which grows quadratically with $K$,
    while $\Delta v$ is a coefficient depending only on the system's spectral structure.

    The \emph{critical point} corresponds to the special case where $v_0$ diverges.  
    The critical parameters are
    \begin{equation}
        \alpha_c = 
        \begin{cases}
            1 & \chi = 1, \\
            \gamma^{-2} & \chi = i,
        \end{cases}
        \qquad
        \beta_c = 
        \begin{cases}
            1 & \chi = 1, \\
            \gamma & \chi = i.
        \end{cases}
    \end{equation}
    At criticality, $\Delta v = 1$ is maximal.  
    Far from criticality, the scaling reduces to
    \begin{equation}
        \Delta v \sim \frac{1}{\gamma^2 \alpha} \quad (\textbf{$\alpha$-form}), 
        \qquad
        \Delta v \sim (1 + \gamma^{-2})\, \beta^2 \quad (\textbf{$\beta$-form}).
    \end{equation}
    Therefore, the results derived in the overdamped case from the analysis of the critical $K_c$ ~\eqref{eq:z_c} remains valid in the general case.
 Achieving a non-trivial non-normal gain of order unity requires $K \gtrsim 1$ at criticality, 
 or $K$ larger than the distance to criticality when the system is far from it.

    \section*{Deterministic Periodic Forcing}

    We now consider the case of periodic forcing, which further confirms that the response kernel \( G(\J, t) \) forms the fundamental basis for all amplification mechanisms.
    Specifically, let the external forcing in \eqref{eq:GLE} be 
    \(
        \g(t) = \sqrt{2\delta}\,\nn \sin(\omega t), \quad \omega > 0,
    \)
    where $\mathbf{g}$ is decomposed along the unique non-normal mode $\mathbf{n}$, which dominates the leading non-normal dynamics (terms scaling as $K^2$, with all remaining terms being $\mathcal{O}(1)$ or smaller). The parameter $\delta$ sets the forcing amplitude.
    
    The appropriate measure of potential amplification is the time-averaged mean-square displacement (MSD) \SM
    \begin{equation} \label{eq:vinf_periodic_def}
        \begin{aligned}
            &v_\infty := \lim_{T \to \infty} \frac{1}{T} \int_0^T \|\x(t)\|_2^2 \, dt 
            = 2\delta \pi \left\| \widehat{G}(\J, \omega)\, \nn \right\|_2^2, \\
            &\text{where} \quad \widehat{G}(\J, \omega) := \frac{1}{\sqrt{2\pi}} \int_0^\infty G(\J, t)\, e^{-i \omega t} dt
        \end{aligned}
    \end{equation}
    denotes the Fourier transform of the kernel evaluated at frequency \(\omega\) \SM.

    As with Gaussian forcing, in the two-dimensional case, the MSD takes the form \eqref{eq:uni_form_var} with
    \begin{equation} \label{eq:dv_sin}
        \Delta v = 
        \begin{cases}
            \dfrac{2}{(\omega^2 - \alpha)^2 + 1 + (\gamma \omega)^2} & \text{\textbf{\(\alpha\)-form})} \\
            \dfrac{2 \beta^2}{(\omega^2 - 1)^2 + \beta^2 + (\gamma \omega)^2} & \text{(\textbf{\(\beta\)-form})}
        \end{cases}
        ,\quad
        K = \sqrt{\frac{\kappa^2 - 1}{2}},
    \end{equation}
    where the value \(\chi = 1\) or \(i\) does not affect \(\Delta v\).

    The normal MSD (i.e., when \(K=0\)) diverges near resonance or  atcritical points given by:
    \begin{subequations}
    \begin{align}
        &\text{(\textbf{\(\alpha\)-form})} \quad
        \begin{cases}
            \alpha_c = \omega^2 \pm 1, \quad \gamma_c = 0 & \chi = 1, \\
            \alpha_c = \omega^2, \quad \gamma_c = \omega^{-1} & \chi = i,
        \end{cases} \\
        &\text{(\textbf{\(\beta\)-form})} \quad
        \begin{cases}
            \beta_c = |\omega^2 - 1|, \quad \gamma_c = 0 & \chi = 1, \\
            \beta_c = \gamma_c, \quad \omega = 1 & \chi = i.
        \end{cases}
    \end{align}
    \end{subequations}

    Similarly to the Gaussian forcing case, \(\Delta v = 1\) at resonance, but this value is not maximal.
    The maximal value becomes \(\Delta v = 1\)  when \(\alpha = \omega^2\) and \(\gamma=0\) in the \(\alpha\)-form,
    and for \(\omega=1\), \(\beta=1\), and \(\gamma=0\) in the \(\beta\)-form.

    Furthermore, as the system moves away from resonance, \(\Delta v\) decreases quadratically with the distance to resonance.
    Hence, achieving a non-normal gain \(\mathcal{G}(K) = K^2 \Delta v\) greater than unity requires  the degree of non-normality \(K\) 
    to exceed the distance to resonance. 
    This is consistent with the conclusion already reached from the overdamped Gaussian analysis and the critical value $K_c$ given in \eqref{eq:z_c}.

    \section*{Unified Amplification Formalism}

  All previous results for the mean-square deviation (MSD) of the response of system \eqref{eq:lin} can be expressed in the unified form \eqref{eq:uni_form_var}, encompassing various types of external forcing--Gaussian noise \eqref{eq:var_form} and Sinusoidal forcing (\ref{eq:uni_form_var}) with \eqref{eq:dv_sin}--as well as different parameter regimes.
    
    In this framework, the \emph{non-normal (amplification) gain} is defined as
    \begin{equation} \label{eq:gain_uni}
        \mathcal{G}(K) := 
        \frac{v_\infty(K)}{v_\infty(K=0)} - 1
        = K^{2} \, \Delta v ,
    \end{equation}
    where $K$ quantifies the degree of non-normality and $\Delta v$ depends solely on the spectrum,
    decaying quadratically with the distance to criticality or resonance.

\begin{figure*}
        \centering
        \includegraphics[width=\textwidth, trim={0 3 0 3}, clip]{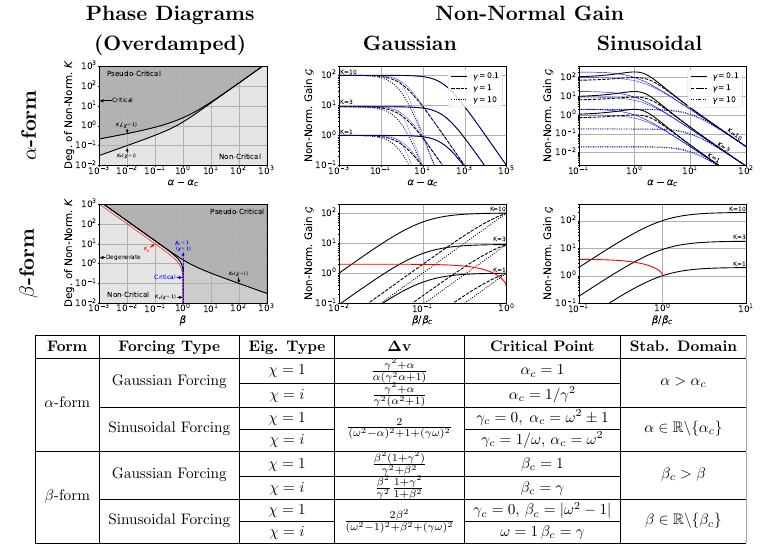}
        \caption{
            \textbf{Phase diagrams} with boundaries between non-critical and pseudo-critical regimes given by \eqref{eq:z_c} and \textbf{unified amplification gain} \eqref{eq:gain_uni} for the mechanisms of spectral criticality, non-normality, and resonance.
            Each \textbf{row} corresponds to one formulation (\textbf{top}: $\alpha$-form; \textbf{bottom}: $\beta$-form). \\
            \textbf{(Left column)}:
            Pseudo-critical phase diagrams for an overdamped system with Gaussian forcing.
            When $K > K_c$, the system is in a pseudo-critical phase where each noise perturbation is transiently amplified before decaying.
            When $K < K_c$, perturbations decay exponentially.
            The threshold $K_c$ is given in \eqref{eq:z_c}.
            For the $\alpha$-form, the critical point is $\alpha_c = 1$ for $\chi = 1$ and $\alpha_c = 0$ for $\chi = i$.
            For the $\beta$-form, the critical point is $\beta_c = 1$ for $\chi = 1$ (shown in blue), while no critical point exists for $\chi = i$.
            In both forms, the corresponding $K_c$ values are indicated for $\chi = 1$ and $\chi = i$.
            Dark grey regions mark the pseudo-critical phase; light grey regions indicate the non-critical phase,
            including degenerate points ($\beta = 0$) and critical points ($\alpha = \alpha_c$, $\beta = \beta_c$). 
            \textbf{(Middle and right columns)}:
            Non-normal gain for two types of forcing (\textbf{middle}: Gaussian; \textbf{right}: Sinusoidal).
            Results are shown for three degrees of non-normality $K = 1, 3, 10$.
            For all $\alpha$-form cases, and for the $\beta$-form with Gaussian forcing, we use three damping values:
            $\gamma = 0.1$ (solid), $\gamma = 1$ (dashed), and $\gamma = 10$ (dotted).
            In the $\alpha$-form, black lines correspond to $\chi = 1$ and thin blue lines to $\chi = i$.
            For the $\beta$-form with Gaussian forcing, only $\chi = 1$ is plotted, since the non-normal gain as a function of $\beta / \beta_c$ is invariant under $\gamma \to 1/\gamma$.
            For the $\beta$-form with Sinusoidal forcing, both $\chi = 1$ and $\chi = i$ produce identical curves as a function of $\beta / \beta_c$ when the system pulsation and damping are set to their critical values. \\
            For the $\beta$-form, the parametrization $(\beta(\epsilon),K(\epsilon))$ \eqref{eq:eps} \cite{Charan2020} is shown in \textbf{red},
            and with $\gamma=1$ for the Gaussian noise non-normal gain figure. \\
            \textbf{(Bottom)}:
            Table listing the coefficient $\Delta v$ (entering in the definition of the non-normal gain~\eqref{eq:gain_uni}),
            as well as the critical point and stability domain for each form and forcing type.
            The overdamped limit can be recovered as the leading order term in power of $\gamma\gg 1$.
        }
        \label{fig:phase_gen}
    \end{figure*}

    Figure \ref{fig:phase_gen} summarizes the unified formalism.  
    For each forcing type (Gaussian or Sinusoidal), the degree of non-normality 
    is determined by the eigenbasis condition number $\kappa$, but with distinct functional forms:
    \begin{equation}
        K = 
        \begin{cases}
            \frac{1}{2}|\kappa - \kappa^{-1}|, & \textbf{(Gaussian)} \\[4pt]
            \sqrt{\frac{\kappa^{2} - 1}{2}}, & \textbf{(Sinusoidal} 
        \end{cases}.
    \end{equation}

    The phase diagrams in Figure \ref{fig:phase_gen} plot $\mathcal{G}(K)$ \eqref{eq:gain_uni}  
    as a function of the distance $\alpha - \alpha_c$ to criticality for the $\alpha$-form,  
    and $\beta / \beta_c$ for the $\beta$-form.  
    These diagrams reveal the values of $K$ needed to achieve various gain level $\mathcal{G}(K)$.

    Two general trends emerge:
    \begin{enumerate}
        \item \textbf{Far from criticality}: As the system moves away from the critical line, larger $K$ values are required to sustain large non-normal gain.
        \item \textbf{Close to criticality}: $\Delta v$ approaches its maximum, so amplification is primarily governed by $K$.
        Conversely, near eigenvalue degeneracy ($\beta = 0$), the non-normal term is suppressed and the gain remains close to its maximal value away from criticality.
    \end{enumerate}

    By unifying criticality, non-normality, and resonance into a single response map,  
    the phase diagrams in Figure \ref{fig:phase_gen} act as ``risk maps''.  
    They identify regions of parameter space where even weak stochastic or periodic perturbations can produce disproportionately large responses.  
    With experimentally estimated pairs $(\beta, K)$ or $(\alpha, K)$, one can:
    \begin{enumerate}
        \item \emph{Predict variance inflation} --
        use $\mathcal{G}(K)$ to quantify the MSD increase relative to the normal ($K=0$) case.
        \item \emph{Diagnose the dominant amplification route} --
        points near criticality or resonance indicate classical mechanisms,
        whereas large $K$ at moderate distances from criticality signal a pseudo-critical regime.
    \end{enumerate}

    \section*{Application to Remote Earthquake Triggering}

    Dynamical triggering refers to the induction of earthquakes by weak,
    low-frequency seismic waves originating from distant large events.
    These waves can provoke new ruptures thousands of kilometers away,
    well beyond typical aftershock zones \cite{Pollitz2012,Mendoza2016,O'Malley2018}.
    Despite extensive study, the phenomenon remains incompletely understood,
    as its causal mechanisms have yet to be definitively identified,
    with several competing hypotheses still unvalidated \cite{Chengzhi_reviewtrig2023}.

    Recently, \cite{Charan2020,Charan2021Remote} proposed a novel mechanism based on the breaking of Hamiltonian symmetry caused by the interplay of rotational effects and friction within fault gouges.
    This mechanism predicts that stressed systems develop extreme sensitivity to small perturbations of any frequency,
    without requiring resonance. Here, we reinterpret the formulation of \cite{Charan2020,Charan2021Remote} within our framework,
    revealing a significantly broader domain of applicability.

    Starting from the general linear dynamical equation \eqref{eq:lin},
    consider the reduced ``normal form'' proposed in \cite{Charan2021Remote}
    \begin{equation}
        \mathbf{J} =
        \begin{pmatrix}
            -1 + \delta & -\eta \\
            \eta & -1 - \delta
        \end{pmatrix}.
    \label{eq:normal_form}
    \end{equation}
    A straightforward spectral decomposition (see \SM) yields
    \begin{equation}
        \lambda_{\pm} = -1 \pm \beta,
        \quad
        \beta = \sqrt{\delta^{2} - \eta^{2}},
        \quad
        \kappa = \left| \sqrt{\frac{\delta + \eta}{\delta - \eta}} \right|.
    \label{eq:pseudo_para}
    \end{equation}
    The critical point is reached as $\eta \to \delta$,
    where the two eigenvalues coalesce at $-1$, signaling a bifurcation in the underdamped system.

    The condition number $\kappa > 1$ diverges as $\eta \to \delta$.
    \cite{Charan2020} introduced the parameter
    \(
        \epsilon := 1 - \frac{\eta}{\delta} > 0,
    \)
    which measures the distance to criticality. For $\epsilon \ll 1$, 
    \(
        \beta \sim \sqrt{\epsilon}.
    \)
    However, this parameterization also implies
    \(
        \kappa \sim \frac{1}{\sqrt{\epsilon}},
    \)
    collapsing the Hopf degeneracy ($\beta \to 0$) and non-normality ($\kappa \to \infty$) into a single control parameter $\epsilon$.
    This obscures which mechanism dominates amplification.

    To disentangle these effects and clarify how criticality, Hopf degeneracy, and non-normality counteract one another along the path defined by this parameterisation, we extract $(\beta, K)$ for each case
    \begin{equation}
        \beta = \delta \sqrt{2 \epsilon - \epsilon^2}, \quad
        K = 
        \begin{cases}
            \displaystyle \frac{2(1-\epsilon)}{\sqrt{2 \epsilon - \epsilon^2}} & \text{(\textbf{Gaussian})} \\
            \displaystyle \sqrt{\frac{1-\epsilon}{\epsilon}} & \text{(\textbf{Solenoidal})}
        \end{cases},
    \label{eq:eps}
    \end{equation}
        where we choose $\delta>\eta$ to ensure operation in the $\beta$-framework with real eigenvalues i.e. $\chi=1$.
        For $\epsilon=0$, the system is degenerate.
        For $\epsilon=1$, the system is critical ($\beta=\delta$).
        We use the parametrization \eqref{eq:eps} to move from degeneracy to criticality by setting  $\delta=\beta_c$.
Depending on the regime, the correspondence can be read from the table in Figure \ref{fig:phase_gen} (bottom).
This parametrization, shown in red in Figure \ref{fig:phase_gen}, satisfies $K \beta \sim \beta_c$, so the divergence of $K$
is exactly compensated by the vanishing of $\beta$ as $\epsilon \to 0$.
The resulting phase diagram and non-normal gain are plotted in Figure \ref{fig:phase_gen}.
Despite the divergence of $K$, the system never reaches pseudo-criticality, and no non-normal amplification occurs.

    For this application, our main contribution is to reveal that while, the parameterization $\epsilon = 1 - \eta/\delta$ naturally monitors the distance to criticality,
    it fails to detect the existence of a ``highly non-normal'' regime ($K > K_c$).
    To address this, we propose re-parameterizing $\mathbf{J}$ as
    \begin{equation}
        \begin{cases}
            \delta = \frac{\beta}{2} \left( \kappa + \kappa^{-1} \right), \\
            \eta = \frac{\beta}{2} \left( \kappa - \kappa^{-1} \right),
        \end{cases}
        \quad \beta \ge 0, \quad \text{for } \delta > \eta.
    \label{eq:re_para}
    \end{equation}
    For $\eta > \delta$, the two expressions are permuted.
    This parameterization allows independent exploration of spectral degeneracy through $\beta$ and non-normality through $\kappa$.
    It also shows that $\delta, \eta > 1$ is permissible as long as $\beta = \sqrt{\delta^2 - \eta^2}$ \eqref{eq:pseudo_para} ensures asymptotic stability.

    Varying $\kappa$ at fixed $\beta$ corresponds to moving vertically in the phase diagram of Figure \ref{fig:phase_gen},
    thereby exploring the full pseudo-critical sector that was previously excluded by the single-parameter $\epsilon$ approach of \cite{Charan2020,Charan2021Remote}.

    Our analysis demonstrates that non-normality provides an independent amplification mechanism,
    capable of producing large responses to weak periodic perturbations
    -- even when the system is spectrally far from resonance or criticality.
    Recognizing non-normality as a separate axis of fault instability significantly expands the hazard landscape:
    faults considered ``sub-critical'' by eigenvalue analysis may nonetheless reside within the pseudo-critical region of Figure \ref{fig:phase_gen},
    where small disturbances can be strongly amplified.

    Incorporating the non-normality parameter $K$ alongside the spectral parameter $\beta$ thus elevates the risk assessment of remote triggering and suggests new diagnostic metrics
    -- such as transient growth bounds or local condition number estimates --
    that could be derived from seismic data.

    \section*{Non-Hermitian Quantum Optics: Pseudo-Critical Regime}

   Non-Hermitian quantum physics has rapidly become a focal point of modern science, not only for its elegant description of open systems with gain and loss, but also for its far-reaching implications in quantum technologies. By extending quantum theory beyond the Hermitian paradigm, it offers new insights into the measurement problem, decoherence, and control of quantum states--key challenges for quantum computing and sensing. Central to this framework are exceptional points (EPs), spectral singularities where eigenvalues and eigenvectors merge, acting as critical organizing centers for wave dynamics and enabling phenomena such as mode switching, unidirectional transport, and extreme sensitivity \cite{Krasnok2021PT_EP,el2018non}.
    Near an EP, small perturbations can induce disproportionately large responses,
    paralleling the spectral criticality analyzed in this work.
    Here, we demonstrate that, for appropriate parameter regimes,
    non-Hermitian quantum optical systems can be recast into our $\alpha$-form framework and exhibit pseudo-critical amplification.

    Non-Hermitian extensions of quantum optics, such as the Forward Four-Wave Mixing (FFWM) experiment \cite{Yue2019},
    can be expressed in the general linear form
    \begin{equation}
        \frac{\partial}{\partial z} \x = \J \x + \g,
        \quad
        \J =
        \begin{pmatrix}
            -a + i \frac{\Delta k}{2} & i c \\
            -i c & l - i \frac{\Delta k}{2}
        \end{pmatrix},
    \label{eq:ffwm}
    \end{equation}
    where $a > 0$ models linear absorption due to imperfect electromagnetically induced transparency (EIT),
    $l$ denotes the small Raman gain from the pump,
    $c$ is the nonlinear coupling coefficient,
    and $\Delta k$ represents the phase mismatch.
    The forcing term $\g$ accounts for injected probe and conjugate fields.
    Although the wave propagates along the spatial coordinate $z$,
    mathematically this is analogous to temporal wave propagation.

    In the FFWM model \eqref{eq:ffwm}, the exceptional point
    -- where eigenvalues and eigenvectors coalesce --
    naturally emerges within the $\beta$-framework \SM.
    At this point, the eigenvalue collision suppresses non-normal amplification,
    mirroring behavior observed in frictional amorphous solids.
    By imposing an appropriate hierarchy among the hyperparameters $(a, l, c, \Delta k)$,
    the system can be mapped onto the $\alpha$-framework, yielding
    \begin{equation}
        \alpha \approx \frac{a - l}{\sqrt{\Delta k^2 - 4 c^2}}, \quad
        \kappa \approx \sqrt{\left| \frac{\Delta k + 2 c}{\Delta k - 2 c} \right|},
    \end{equation}
    which identifies the parameter regime
    \begin{equation}
        \Delta k \approx 2 c \gg |\Delta k - 2 c| \gg a + l > a - l > 0,
    \end{equation}
    under which pseudo-critical amplification arises despite the presence of optical damping.
    Here, $\alpha$ quantifies the distance to spectral criticality,
    while the pair $(\alpha, \kappa)$ jointly govern the interplay between criticality and non-normality.

    Since $\mathbf{x}$ represents the slowly-varying field amplitudes, its norm squared,
    $\|\mathbf{x}\|^2$, is directly proportional to the optical energy.
    In the pseudo-critical regime ($K > K_c(\alpha)$, see \eqref{eq:z_c}),
    transient growth can occur: the wave transiently extracts net energy from the medium despite friction-like optical damping before ultimately decaying.
    This counterintuitive gain-from-loss effect is a direct manifestation of non-normal amplification in a propagating-wave context.
 Thus, by tuning FFWM parameters into the $\alpha$-form pseudo-critical regime,
    it is possible to engineer transient optical energy amplification without requiring true lasing or exact $\mathcal{PT}$-symmetry breaking.

    \section*{Conclusion}

   We have presented a unified linear framework that both distinguishes and quantitatively links two key amplification mechanisms:
    conventional criticality/resonance (governed by the spectral gap) and pseudo-criticality (driven by non-normality, measured by the condition number $\kappa$ or degree of non-normality $K$).
    For both Gaussian noise and sinusoidal forcing, we have derived closed-form expressions for the mean-square deviation (MSD), showing that these mechanisms contribute together.
    This yields a universal master law visualized in phase diagrams (Figure \ref{fig:phase_gen}),
    where any linear system can be placed and its susceptibility to large deviations assessed.

    Reanalyzing the ``giant amplification'' model of remote earthquake triggering from \cite{Charan2020,Charan2021Remote}, we showed that its single-parameter setup constrains the system below the pseudo-critical threshold $K=K_c$. Our two-parameter reformulation lifts this limit, allowing $\kappa$ to grow independently of $\lambda$ and thus accessing the full non-normal regime which has a much broader domain of existence.

    The same analysis applies to parity-time symmetry breaking in non-Hermitian physics.
    In particular, the four-wave mixing in cold atoms \cite{Yue2019} corresponds to an overdamped Langevin framework,
    with a linear dynamics matrix matching \eqref{eq:normal_form} up to a $\pi/2$ phase shift.
    Thus, our conclusions extend to exceptional points in non-Hermitian systems,
    where an emphasis on bifurcation and critical phenomena often obscures the role of non-normality.

    The proposed framework is universal, applying to systems from seismology to non-Hermitian photonics and ecological networks.
    A nonlinear extension suggests the coupling between the two key amplification mechanisms persists beyond linearity.
    This opens new diagnostic paths: 
    (\emph{i}) in seismology, transient growth bounds and field estimates of $\kappa$ could complement traditional eigenvalue-based hazard metrics;
    and (\emph{ii}) in laboratory or numerical studies of exceptional points,
    the phase diagrams developed here give a straightforward recipe for separating genuine critical point and degeneracy effects from non-normal transients.
    More broadly, any discipline that diagnoses ``critical-like'' bursts should test whether they originate from criticality,
    large $K$, or the potential combination of both
    -- a distinction now made precise by the framework laid out in this paper.

  
    \bibliographystyle{naturemag}  
    \bibliography{bibliography} 


    \clearpage
    \onecolumn
    
    \centerline{\bf \Large Supplementary Materials}
    \vspace{2cm}

    \appendix

    \tableofcontents
    \newpage

    \section{Reduced Non-Normal Forms}

    In this work, we consider the following generic \(N\)-dimensional model:
    \begin{equation}
        \label{eq:apx_langevin}
        \ddot{\x} + \gamma \dot{\x} = \f(\x) + \g(t),
    \end{equation}
    where \(\g(t)\) is an arbitrary time-dependent forcing term and \(\f(\x)\) is a (not necessarily variational) vector field.  
    We denote by
    \(
        \J = \left. \D \f(\x) \right|_{\x = \OO}
    \)
    the Jacobian of \(\f\) at the equilibrium \(\x^{*} = \OO\),
    chosen as the origin without loss of generality.  
    Throughout, we assume that \(\J\) is negative-definite, diagonalizable, and non-normal.

    Recall that a matrix \(\J\) is \emph{non-normal} if \([\J, \J^\dagger] \neq 0\),
    or equivalently, if it cannot be unitarily diagonalized.
    Non-normality is well known to give rise to transient growth and related temporary instabilities.

    \subsection{Non-Normal $\alpha$-Form}

    It was shown in \cite{troude2024} that any \emph{highly} non-normal linear system of dimension \(N\) can
    -- after a suitable unitary change of basis and time rescaling --
    be reduced to the \(2\times 2\) family
    \begin{equation}
        \label{eq:apx_reduce_nn_alpha}
        \Gam_{\chi}^\alpha
        =
        \begin{pmatrix}
            -\alpha & \chi \kappa \\
            \chi \kappa^{-1} & -\alpha
        \end{pmatrix},
        \quad \chi \in \{1, i\}.
    \end{equation}

    Consider the overdamped system
    \begin{equation}
        \dot{\x} = \Gam^\alpha _{\chi} \, \x + \g(t).
    \end{equation}
    When \(\chi = 1\), the eigenvalues are \(\lambda_{\pm} = -\alpha \pm 1\),
    so the system approaches a real-eigenvalue bifurcation as \(\alpha \to 1^+\).  
    For \(\chi = i\), one has \(\lambda_{\pm} = -\alpha \pm i\),
    and a Hopf bifurcation occurs when \(\alpha \to 0^+\).

    Such a representation can be obtained by considering a \(2\times 2\) diagonalizable matrix \(\J\),
    such that \(\J = \PP \Lamb \PP^{-1}\), where \(\PP\) is the eigenbasis transformation matrix.
    We write \(\PP = (\pp_{+}, \pp_{-})\), where \(\pp_{\pm}\) are the normalized right eigenvectors of \(\J\).

    In \cite{troude2024}, the passage from \(\J\) to \(\Gam_\chi^\alpha\) in \eqref{eq:apx_reduce_nn_alpha} is obtained by explicitly tracking the non-orthogonality of the eigenvectors \(\pp_{\pm}\).
    The more the eigenvectors coalesce, the more \(\J\) is non-normal,
    which is quantified by the condition number \(\kappa\) of \(\PP\),
    defined as the ratio between the largest and smallest singular values of \(\PP\).

    Since we are considering a \(2\)-dimensional case
    -- minimal for exhibiting non-normal behaviour -- 
    we can explicitly compute the singular value decomposition (SVD) of \(\PP\):  
    \(\PP = \U \Sig \V^\dag\), where \(\U\) and \(\V\) are unitary matrices,
    and \(\Sig = \text{Diag}(\sigma_{+}, \sigma_{-})\) is the diagonal matrix of singular values.  
    From \(\PP = (\pp_{+}, \pp_{-})\), with normalized columns, we have
    \begin{equation}
        \PP^\dag \PP = 
        \V \Sig^2 \V^\dag = 
        \begin{pmatrix}
            1 & c \\ c^* & 1
        \end{pmatrix},
        \quad \text{where} \quad c = \pp_{+}^\dag \pp_{-}.
    \end{equation}
    The singular values of \(\PP\) and the matrix \(\V\) are
    \begin{equation}
        \sigma_{\pm} = 1 \pm |c|,
        \quad
        \V = \frac{1}{\sqrt{2}}
        \begin{pmatrix}
            e^{i\phi} & -e^{i\phi} \\
            1 & 1
        \end{pmatrix},
        \quad \text{where} \quad \phi = \arg c.
    \end{equation}

    Let the eigenvalues of \(\J\) associated to \(\pp_{\pm}\) be \(\lambda_{\pm} = \lambda \mp \chi \Delta\),
    where \(\Delta > 0\) and \(\chi = 1\) if the eigenvalues are real,
    \(\chi = i\) if they are complex conjugates.  
    Up to a unitary transformation:
    \begin{align}
        \J &= \PP \Lamb \PP^{-1} 
        = \U \Sig \V^\dag \Lamb \V \Sig^{-1} \U^\dag \\
        &= \frac{1}{2} \U \Sig
        \begin{pmatrix}
            e^{-i\phi} & 1 \\
            -e^{-i\phi} & 1
        \end{pmatrix}
        \begin{pmatrix}
            \lambda_{+} & 0 \\ 0 & \lambda_{-}
        \end{pmatrix}
        \begin{pmatrix}
            e^{i\phi} & -e^{i\phi} \\
            1 & 1
        \end{pmatrix}
        \Sig^{-1} \U^\dag \\
        &= \U
        \begin{pmatrix}
            \sigma_{+} & 0 \\ 0 & \sigma_{-}
        \end{pmatrix}
        \begin{pmatrix}
            \lambda & \chi \Delta \\
            \chi \Delta & \lambda
        \end{pmatrix}
        \begin{pmatrix}
            \sigma_{+} & 0 \\ 0 & \sigma_{-}
        \end{pmatrix}^{-1}
        \U^\dag .
    \end{align}

    We therefore define
    \begin{equation}
        \label{eq:apx_Gam}
        \Gam_{\chi} = \U^\dag \J \U =
        \begin{pmatrix}
            \lambda & \chi \Delta \kappa \\
            \chi \Delta \kappa^{-1} & \lambda
        \end{pmatrix}.
    \end{equation}
    Thus, up to a unitary transformation, \(\kappa\) quantifies the non-normality of \(\J\),
    while the pair \((\lambda, \Delta)\) describes the criticality for each scenario \(\chi \in \{1, i\}\).

    Following \cite{troude2024}, we rescale time via \(t \to t \Delta\), giving
    \begin{equation}
    \Gam_{\chi}^\alpha = \frac{1}{\Delta} \Gam_{\chi} =
    \begin{pmatrix}
        -\alpha & \chi \kappa \\
        \chi \kappa^{-1} & -\alpha
    \end{pmatrix},
    \quad \text{where} \quad \alpha = -\frac{\lambda}{\Delta}.
    \end{equation}

    This \(\alpha\)-form is especially useful for studying the relationship between non-normality and criticality in overdamped systems:  
    for \(\chi = 1\), the system is critical as \(\alpha \to 1^+\); for \(\chi = i\), it is critical as \(\alpha \to 0^+\).
    In both cases, the system becomes asymptotically far from criticality away from these limits.
    The reduced non-normal \(\alpha\)-form therefore enables simultaneous quantification of non-normality and criticality in overdamped systems.
    \newline

    For an underdamped system
    \begin{equation}
    \ddot{\x} = \J \x + \g(t),
    \end{equation}
    the eigenvalues are \(\pm \sqrt{\lambda_{\pm}}\), where \(\lambda_{\pm}\) are the eigenvalues of \(\J\). In the \(\alpha\)-form framework, these take the form \(\pm i \sqrt{\alpha \pm \chi}\).  

    If \(\chi = 1\), stability requires \(\alpha \pm 1 \ge 0\), so \(\alpha \to -1^{+}\) defines a Hamiltonian–Hopf bifurcation, occurring at the same point as in the overdamped case.  
    If \(\chi = i\), the system is always unstable; transitioning from \(\chi = 1\) to \(\chi = -1\) corresponds to a Hopf bifurcation.

    Although the \(\alpha\)-form is natural and convenient for overdamped cases, it does not allow for a continuous transition between the two cases \(\chi = 1\) and \(\chi = i\).  
    In the next section, we introduce a second non-normal form that enables such a continuous transition, making it more suitable for studying the general system \eqref{eq:apx_langevin}.

    \subsection{Non-Normal $\beta$-Form}

    In \eqref{eq:apx_Gam}, rather than introducing the time rescaling \(t \to t / \Delta\),
    where \(\Delta = |\lambda_{+} - \lambda_{-}| / 2\) and \(\lambda_{\pm}\) are the eigenvalues of \(\J\),  
    to obtain the \(\alpha\)-form \eqref{eq:apx_reduce_nn_alpha},  
    we may instead consider the rescaling \(t \to t / |\lambda|\),  
    where \(\lambda = (\lambda_{+} + \lambda_{-}) / 2 < 0\).  
    If \(\lambda > 0\), the system is never stable.

    While the change of sign of \(\lambda\) (inducing a bifurcation) is already embedded in the \(\alpha\)-form,  
    here we wish to capture the analogous behaviour when \(\Delta \to 0\).  
    This leads us to define the \(\beta\)-form:
    \begin{equation}
        \Gam_{\chi}^\beta = \frac{1}{|\lambda|} \Gam_{\chi} =
        \begin{pmatrix}
            -1 & \chi \beta \kappa \\
            \chi \beta \kappa^{-1} & -1
        \end{pmatrix},
        \quad \text{where} \quad
        \beta = \frac{\Delta}{|\lambda|} > 0.
    \end{equation}

    In this representation, the eigenvalues of \(\Gam_{\chi}^\beta\) are \(-1 \pm \beta \chi\) (\(\chi = 1, i\)),  
    and the eigenvalues of the corresponding underdamped system are \(\pm i \sqrt{\,1 \pm \beta \chi\,}\).  
\begin{itemize}
    \item For \(\chi = 1\), a Hamiltonian-Hopf bifurcation occurs as \(\beta \to 1^{-}\),  
    which coincides with the critical point in the overdamped case.  
    \item For both \(\chi = 1\) and \(\chi = i\), \(\beta = 0\) represents a degenerate point where all eigenvalues are equal (\(\pm i\)),  
    allowing a continuous transition between the two cases.
\end{itemize}

    If the system is overdamped, the transition at \(\beta = 0\) does not induce criticality
    -- the system remains stable --
    but it does enable a smooth transition from a monotonically decaying regime to a damped oscillatory regime.

    \subsection{Summary: Reduced Non-Normal Forms}

    Any non-normal system can be represented, up to a unitary transformation,
    by one of the following two \(2 \times 2\) reduced forms
    \begin{align}
        &\textbf{\(\alpha\)-form} \qquad
        \Gam_{\chi}^\alpha =
        \begin{pmatrix}
            -\alpha & \chi \kappa \\[4pt]
            \chi \kappa^{-1} & -\alpha
        \end{pmatrix}, \\[6pt]
        &\textbf{\(\beta\)-form} \qquad
        \Gam_{\chi}^\beta =
        \begin{pmatrix}
            -1 & \chi \beta \kappa \\[4pt]
            \chi \beta \kappa^{-1} & -1
        \end{pmatrix}.
    \end{align}
    These forms encompass all bifurcations of overdamped and underdamped systems when combined with non-normal amplification.  

    In our subsequent analytical and numerical studies,
    we work with these reduced non-normal forms, as they capture both:
    \begin{itemize}
        \item the approach to criticality, \(\mathrm{Re}(\lambda_{+}) \to 0^{-}\), and
        \item the pseudo-critical growth driven by \(\kappa \to \infty\).
    \end{itemize}

    \section{General Derivation of the Kernel for a System with a Unique Non-Normal Mode}

    Ref. \cite{troude2025} showed that, if a matrix \(\J\) possesses a \emph{unique non-normal mode},  
    its eigenbasis transformation \(\PP\) can be expressed (up to a unitary transformation) as
    \(
        \PP = \I + \left( \kappa^{-1} - 1 \right) \nn \nn^\dag,
    \)
    where \(\kappa\) is the condition number of \(\PP\) (thus quantifying the degree of non-normality of \(\J\)),  
    and \(\nn\) is the unique non-normal mode of \(\J\).

    This framework is useful because it allows one to explicitly determine the dependence of matrices, such as \(G(\J, t)\),  
    where \(G(\cdot, t)\) is a smooth matrix-valued function and \(t \in \mathbb{R}\) represents time,
    as a function of the condition number \(\kappa\) of \(\PP\).
    In this context, \(G(\J, t)\) can be interpreted as a \emph{kernel matrix}.

    We proceed as follows:
    \begin{enumerate}
        \item Show how \(\kappa\) influences the quadratic variation of a Gaussian dynamical process under a kernel \(G(\J, t)\).
        \item Present an analogous result for the average squared deviation in a system with sinusoidal forcing.
        \item Derive the explicit form of the kernel \(G(\J, t)\) for a linear second-order ODE with isotropic friction and time-dependent forcing.
        \item Apply the non-normal mode framework to a kernel-structured system, obtaining explicit expressions in the two \(\alpha/\beta\)-formalisms.
    \end{enumerate}

    \subsection{Non-Normal Kernel}

    We study here the impact of non-normality on a stable \(N\)-dimensional dynamical system of the form
    \begin{equation}
    \x_t ~=~ \int_{0}^{t} G(\J, t - s) \, \g(s) \, ds,
    \end{equation}
    where \(G(\cdot, t)\) is a smooth kernel function, \(\J\) is a non-normal,
    diagonalizable matrix, and \(\g(t)\) is an external time-dependent forcing term.

    We consider two types of forcing:  
    a \textbf{stochastic} one given by Gaussian white noise, and a \textbf{deterministic periodic} one given by a solenoidal function:
    \begin{equation}
        \g(t) =
        \begin{cases}
        \sqrt{2\delta} \, \et_t, 
        \quad \et_t \sim \mathcal{N}(\OO, \I), 
        \quad \mw{\et_t \et_s^\dag} = \delta(t - s) \, \I,
        & \textbf{(Stochastic)}, \\[6pt]
        \sqrt{2\delta} \, \hat{\g} \, \sin(\omega t),
        & \textbf{(Periodic)},
        \end{cases}
    \end{equation}
    where:
    \begin{itemize}
        \item \(\delta\) sets the noise intensity (stochastic case) or the forcing amplitude (periodic case),
        \item \(\mw{\cdot} := \mathbb{E}[\cdot]\) denotes the expectation over realizations,
        \item \(\omega\) is the forcing frequency,
        \item \(\hat{\g}\) is a normalized forcing direction.
    \end{itemize}

    In both cases, we assume that \(\J\) admits the spectral decomposition
    \(
    \J = \PP \Lamb \PP^{-1}, \;
    \Lamb = \mathrm{Diag}(\lambda_{1}, \dots, \lambda_{N}),
    \)
    where the \(\lambda_{i}\) are the eigenvalues of \(\J\), and
    \(
    \PP = \I + \left( \kappa^{-1} - 1 \right) \nn \nn^\dag,
    \)
    with \(\nn\) a normalized vector representing the unique non-normal mode,  
    and \(\kappa\) the condition number of \(\PP\), quantifying the degree of non-normality.

    Our goal is to show how the non-normality of \(\J\) (as measured by  \(\kappa\)) amplifies the system's response.  
    To do this, we define in both cases a measure of \emph{variance} or \emph{average squared amplitude}:
    \begin{equation}
        \label{eq:apx_square_dev}
        v_{\infty} =
        \begin{cases}
        \displaystyle
        \lim_{t \to \infty} \, \mw{\|\x_t\|^{2}} 
        = 2\delta \int_{0}^{\infty} \| G(\J, t) \|^{2} \, dt,
        & \textbf{(Stochastic)}, \\[12pt]
        \displaystyle
        \lim_{t \to \infty} \, \frac{2\delta}{t} \int_{0}^{t} \|\x_s\|^{2} \, ds,
        & \textbf{(Periodic)}.
        \end{cases}
    \end{equation}

    The objective of this section is to determine the explicit dependence of \(v_{\infty}\) on \(\kappa\)  
    in both the stochastic and deterministic periodic cases.  
    We begin with the stochastic forcing case, and then derive the corresponding expression for the periodic forcing.

    \subsubsection{Non-Normal Kernel for Gaussian Processes}
    \label{apx:non_normal_kernel}

    Consider \(N\) independent Gaussian processes \(\et_t\) at each time \(t\),  
    such that \(\mw{\et_t \et_s^\dag} = \delta(t-s) \, \I\).  
    The process \(\x_t\) is given by
    \begin{equation}
        \x_t ~=~ \sqrt{2\delta} \, \int_{0}^{t} G(\J, t-s) \, \et_s \, ds,
    \end{equation}
    where \(G(\cdot, t)\) is a smooth matrix-valued function and \(\J\) is a square, diagonalizable matrix  
    satisfying \(\int_{0}^{t} G(\J, s) \, G(\J, s)^\dag \, ds < \infty\) as \(t \to \infty\).

    The variance of \(\x_t\) is
    \begin{equation}
        \mw{\|\x_t\|_2^2} = 2\delta \int_{0}^{t} \|G(\J, s)\|_F^2 \, ds,
    \end{equation}
    where \(\|A\|_F = \sqrt{\mathrm{Tr}(A A^\dag)}\) denotes the Frobenius norm.  

    We assume that \(\J\) is diagonalizable with eigenbasis transformation
    \(
        \PP = \I + (\kappa^{-1} - 1) \, \nn \nn^\dag,
    \)
    where \(\kappa\) is the condition number of \(\PP\) (measuring the degree of non-normality)  
    and \(\nn\) is the unique non-normal mode.  
    If \(\J = \PP \Lamb \PP^{-1}\) is the spectral decomposition of \(\J\), then
    \(
    G(\J, t) = \PP \, G(\Lamb, t) \, \PP^{-1}.
    \)
    We can then express \(\|G(\J, s)\|_F^2\) in terms of \(\kappa\) as follows:
    \begin{align}
        \|G(\J, s)\|_F^2
        &= \mathrm{Tr}\!\left( G(\J, t) \, G(\J, t)^\dag \right) \\
        &= \mathrm{Tr}\!\left( \PP G(\Lamb, t) \PP^{-1} (\PP^{-1})^\dag G(\Lamb, t)^\dag \PP^\dag \right) \\
        &= \mathrm{Tr}\!\left( G(\Lamb, t) (\PP^\dag \PP)^{-1} G(\Lamb, t)^\dag (\PP^\dag \PP) \right) \\
        &= \mathrm{Tr}\!\left( G(\Lamb, t) \left[ \I + (\kappa^2 - 1) \nn \nn^\dag \right]
                            G(\Lamb, t)^\dag \left[ \I + (\kappa^{-2} - 1) \nn \nn^\dag \right] \right) \\
        &= \|G(\Lamb, t)\|_F^2
        + (\kappa - \kappa^{-1})^2
            \left[
                \mathrm{Tr}\!\left( \nn \nn^\dag |G(\Lamb, t)|^2 \right)
                - \mathrm{Tr}\!\left( \nn \nn^\dag G(\Lamb, t)^\dag \nn \nn^\dag G(\Lamb, t) \nn \nn^\dag \right)
            \right] \\
        &= \|G(\Lamb, t)\|_F^2
        + (\kappa - \kappa^{-1})^2
            \left[
                \mw{|G(\Lamb, t)|^2}_n - \left| \mw{G(\Lamb, t)}_n \right|^2
            \right],
        \label{eq:apx_non_normal_kernel_res_gauss}
    \end{align}
    where \(\mw{A}_n := \nn^\dag A \nn\).  
    In deriving this expression, we used the cyclic property of the trace and the fact that  
    \(\nn \nn^\dag = (\nn \nn^\dag)^2\).

    Assume \(G(\lambda, \cdot)\) is strictly decreasing and convex,
    with \(G(\lambda, 0) = 1\) and \(G(\lambda, t) \to 0\) as \(t \to \infty\).  
    Then \(\|G(\Lamb, s)\|_F^2\) decreases monotonically from 1 to 0.  
    However, the factor
    \(
        \mw{|G(\Lamb, t)|^2}_n - \left|\mw{G(\Lamb, t)}_n\right|^2
    \)
    multiplying the non-normality term \((\kappa - \kappa^{-1})^2\) in  
    \eqref{eq:apx_non_normal_kernel_res_gauss} need not be strictly monotonic.

    In general, the squared deviation \eqref{eq:apx_square_dev} for a Gaussian forcing can be decomposed as:
    \begin{equation}
        \label{eq:apx_var_g}
        v_\infty = v_0 \left[ 1 + K^2 \Delta v \right],
    \end{equation}
    where \(v_0\) is the squared deviation for a normal system (\(\kappa = 1\)):
    \begin{equation}
        \label{eq:apx_v0_g}
        v_0 = 2\delta \int_{0}^{\infty} \|G(\Lamb, s)\|_F^2 \, ds,
    \end{equation}
    and
    \begin{equation}
        \label{eq:apx_K_w_g}
        K = \frac{1}{2} \, |\kappa - \kappa^{-1}|, \quad
        \Delta v = \frac{4}{v_0} \int_{0}^{\infty}
        \left[ \mw{|G(\Lamb, t)|^2}_n - \left| \mw{G(\Lamb, t)}_n \right|^2 \right] \, dt.
    \end{equation}
    Thus \(K\) provides an alternative measure of the degree of non-normality.

    In the special case of a two-dimensional system with eigenvalues \(\lambda_\pm\)  
    and non-normal mode \(\n = (1,\,1) / \sqrt{2}\), the Frobenius norm reduces to:
    \begin{equation}    \label{eq:apx_2d_frob}
        \|G(\J, t)\|_F^2
        = |G(\lambda_+, t)|^2 + |G(\lambda_-, t)|^2
        + K^2 \, |G(\lambda_+, t) - G(\lambda_-, t)|^2.
    \end{equation}
    This compact expression is particularly convenient for explicitly computing \(v_0\) and \(\Delta v\)  
    in the reduced non-normal form framework.  
    Moreover, in the degenerate limit where the eigenvalues coincide,  
    the non-normal contribution vanishes.

    \subsubsection{Non-Normal Amplification with a Sinusoidal Source}
    \label{apx:non_normal_sinusoidal}

    We now consider a periodic forcing
    \(
        \g(t) = \sqrt{2\delta} \, \hat{\g} \, \sin(\omega t),
        \; \|\hat{\g}\| = 1,
    \)
    so that the dynamics is
    \begin{equation}
        \x_t = \sqrt{2\delta} \int_{0}^{t} G(\J, t-s) \, \hat{\g} \, \sin(\omega s) \, ds.
    \end{equation}
    For convenience, define the \emph{vector kernel}
    \(
        \mathbf{G}(t) := G(\J, t) \, \hat{\g}.
    \)
    We aim to compute the squared deviation \eqref{eq:apx_square_dev} in the long-time limit.
    \newline

    \paragraph{Fourier decomposition}
    For each component \(x_i\) of \(\x\), write
    \begin{align}
        x_{i,t}
        &= \sqrt{2\delta} \, \mathrm{Im} \left[ e^{i\omega t} \int_{0}^{t} G_i(s) e^{-i\omega s} \, ds \right] \\
        &= 2\sqrt{\delta\pi} \, \mathrm{Im} \left[ e^{i\omega t} \, \hat{G}_i(\omega) \right] - \Delta_t,
        \quad \text{where} \quad
        \Delta_t = \mathrm{Im} \left[ e^{i\omega t} \int_{t}^{\infty} G_i(s) e^{-i\omega s} \, ds \right] \\
        &= 2\sqrt{\delta\pi} \, |\hat{G}_i(\omega)| \, \mathrm{Im} \left[ e^{i(\omega t + \phi)} \right] - \Delta_t,
        \quad \phi := \arg \hat{G}_i(\omega) \\
        &= 2\sqrt{\delta\pi} \, |\hat{G}_i(\omega)| \, \sin(\omega t + \phi) - \Delta_t,
        \label{eq:apx_fourier_xi}
    \end{align}
    where
    \begin{equation}
        \hat{G}_i(\omega)
        = \frac{1}{\sqrt{2\pi}} \int_{0}^{\infty} G_i(s) e^{-i\omega s} \, ds
    \end{equation}
    is the Fourier transform of \(G_i\).  
    The residual term \(\Delta_t\) decays exponentially as \(t \to \infty\) and can be neglected in long-time averages.
    \newline

    \paragraph{Long-time variance}
    From \eqref{eq:apx_fourier_xi}, the variance of each component is
    \begin{equation}
    \lim_{t \to \infty} \frac{1}{t} \int_{0}^{t} |x_{i,s}|^2 \, ds
    = 2\delta\pi \, |\hat{G}_i(\omega)|^2.
    \end{equation}
    Summing over components yields
    \begin{equation}
    v_\infty
    = 2\delta\pi \, \|\hat{G}(\J, \omega) \, \hat{\g}\|_2^2.
    \end{equation}
    \newline

    \paragraph{Effect of a unique non-normal mode}
    Assume \(\J\) is non-normal with a unique non-normal mode \(\nn\),  
    and eigenbasis transformation
    \[
    \PP = \I + (\kappa^{-1} - 1) \, \nn \nn^\dag,
    \]
    so that \(\Lamb = \PP^{-1} \J \PP\).  
    If the forcing is aligned with the non-normal mode, \(\hat{\g} = \nn\), then
    \begin{align}
    \|\hat{G}(\J, \omega) \, \hat{\g}\|_2^2
    &= \|\hat{G}(\J, \omega) \, \nn\|_2^2
    = \nn^\dag \hat{G}(\J, \omega)^\dag \hat{G}(\J, \omega) \nn \\
    &= \kappa^2 \, \nn^\dag \hat{G}(\Lamb, \omega)^\dag (\PP^\dag \PP) \, \hat{G}(\Lamb, \omega) \, \nn
    \quad \text{since} \quad \PP^{-1} \nn = \kappa \, \nn \\
    &= \kappa^2 \left[ \mw{|\hat{G}(\Lamb, \omega)|^2}_n
        + (\kappa^{-2} - 1) \, \left| \mw{\hat{G}(\Lamb, \omega)}_n \right|^2 \right]
    \quad \text{since} \quad \PP^\dag \PP = \I + (\kappa^{-2} - 1) \nn \nn^\dag \\
    &= \left| \mw{\hat{G}(\Lamb, \omega)}_n \right|^2
    + \kappa^2 \left[ \mw{|\hat{G}(\Lamb, \omega)|^2}_n
                    - \left| \mw{\hat{G}(\Lamb, \omega)}_n \right|^2 \right],
    \end{align}
    where \(\mw{\cdot}_n := \nn^\dag (\cdot) \, \nn\).
    \newline

    \paragraph{Two-dimensional case}
    If \(\J\) is \(2\times 2\) with \(\nn = (1,\,1) / \sqrt{2}\) and eigenvalues \(\lambda_\pm\), then
    \begin{equation}
    v_\infty
    = \delta\pi \left[
        |\hat{G}(\lambda_+, \omega)|^2
        + |\hat{G}(\lambda_-, \omega)|^2
        + \frac{\kappa^2 - 1}{2} \, |\hat{G}(\lambda_+, \omega) - \hat{G}(\lambda_-, \omega)|^2
    \right].
    \end{equation}
    \newline

    \paragraph{General expression}
    Analogously to the Gaussian case \eqref{eq:apx_var_g}, we can write
    \begin{equation}
        \label{eq:apx_var_s}
        v_\infty = v_0 \left[ 1 + K^2 \Delta v \right],
        \end{equation}
        where the normal-case variance (\(\kappa = 1\)) is
        \begin{equation}
        v_0 = \delta\pi \left[
            |\hat{G}(\lambda_+, \omega)|^2
            + |\hat{G}(\lambda_-, \omega)|^2
        \right],
    \end{equation}
    and
    \begin{equation}
        K = \sqrt{\frac{\kappa^2 - 1}{2}}, \quad
        \Delta v = \frac{|\hat{G}(\lambda_+, \omega) - \hat{G}(\lambda_-, \omega)|^2}{v_0}.
    \end{equation}
    As in the Gaussian forcing case, \(\Delta v \to 0\) in the degenerate limit  
    (where the eigenvalues coincide), and the non-normal contribution disappears.

    \subsection{Application to Linear $2^\text{nd}$-Order ODE with Time-Dependent Forcing}
    \label{apx:linear_langevin}

    We now apply the preceding general results for smooth kernels to the specific case of a linear, second-order system with isotropic damping and time-dependent forcing:
    \begin{equation}
    \label{eq:apx_langevin_linear}
    \ddot{\x} + \gamma \, \dot{\x} = \J \, \x + \g(t),
    \end{equation}
    where:
    \begin{itemize}
        \item $\g(t)$ is a generic time-dependent forcing term (stochastic or periodic);
        \item $\gamma > 0$ is a scalar damping coefficient;
        \item $\J$ is the Jacobian matrix of the system, evaluated at the equilibrium point $\x^* = \mathbf{0}$,
        assumed diagonalizable, negative-definite, and possibly non-normal.
    \end{itemize}

    The focus is on quantifying how the \emph{non-normality} of $\J$ modifies the squared deviation of the particular solution driven by $\g(t)$.  
    Specifically, we are interested in the long-time limit of the variance
    \(
        v_\infty
    \)
    as defined in \eqref{eq:apx_square_dev}, and how it scales with the condition number $\kappa$ of the eigenbasis transformation $\PP$ of $\J$.  

    In the following, we will:
    \begin{enumerate}
        \item derive the explicit kernel $G(\J,t)$ for \eqref{eq:apx_langevin_linear};
        \item compute $v_0$, $K$, and $\Delta v$ in closed form for both Gaussian and sinusoidal forcing,
        using the reduced non-normal $\alpha/\beta$-forms introduced earlier;
        \item compare the scaling with $\kappa$ to the generic results of Sections~\ref{apx:non_normal_kernel} and~\ref{apx:non_normal_sinusoidal}.
    \end{enumerate}

    \subsubsection{Kernel Derivation}

    To facilitate the analysis, we represent the system state as $\z = (\dot{\x}, \x)$,
    which allows us to rewrite the system dynamics as a set of $2N$ first-order differential equations
    \begin{equation}
        \label{eq:linear_first_order}
        \dot{\z} = \A \z + \gsim,
        \quad \text{where} \quad
        \A =
        \begin{pmatrix}
            -\gamma \I & \J \\
            \I & \OO
        \end{pmatrix},
        \quad
        \gsim =
        \begin{pmatrix}
            \g \\
            \OO
        \end{pmatrix}.
    \end{equation}
    The general solution to this system is
    \begin{equation}
        \label{eq:linear_sol}
        \z_t = e^{\A t} \z_0 + \int_0^t e^{\A (t-s)} \gsim_s \, ds.
    \end{equation}
    In the special case of homogeneous damping in all directions,
    the solution retains the same structure as the overdamped case, but with doubled dimensionality \cite{troude2025}.

    Given the spectral decomposition of $\J$, i.e., $\J = \PP \Lamb \PP^{-1}$,
    we seek a generic solution of \eqref{eq:apx_langevin_linear} expressed in terms of $\J$.

    We begin by computing the characteristic polynomial of $\A$:
    \begin{equation}
        \chi(\mu) = \det(\A - \mu \I) = \det\big(\mu(\mu + \gamma) \I - \J\big).
    \end{equation}
    Thus, for each eigenvalue $\lambda \in \sig(\J)$,
    the eigenvalues $\mu$ of $\A$ satisfy the quadratic equation
    \begin{equation}
        \mu^2 + \gamma \mu - \lambda = 0,
        \quad \Rightarrow \quad
        \mu_\pm(\lambda) = -\frac{\gamma}{2} \pm \sqrt{\left(\frac{\gamma}{2}\right)^2 + \lambda}.
    \end{equation}
    Hence, the spectrum of $\A$ is
    $\sig(\A) = \sig\big(\mu_+(\J)\big) \cup \sig\big(\mu_-(\J)\big)$.

    Next, consider the normalized eigenvector $\pp^A_\mu$ of $\A$ associated to eigenvalue $\mu \in \sig(\A)$.
    Decompose it into two $N$-dimensional blocks as $\pp^A_\mu = (\pp^{A1}_\mu, \pp^{A2}_\mu)$.
    From
    \begin{equation}
        (\A - \mu \I) \pp^A_\mu = \mathbf{0},
    \end{equation}
    we obtain the system
    \begin{equation}
        \begin{cases}
            \J \pp^{A2}_\mu - (\mu + \gamma) \pp^{A1}_\mu = \mathbf{0}, \\
            \pp^{A1}_\mu = -\mu \pp^{A2}_\mu,
        \end{cases}
        \quad \Rightarrow \quad
        \big(\J + \mu(\mu + \gamma) \I \big) \pp^{A2}_\mu = \mathbf{0}.
    \end{equation}
    Thus, $\pp^{A2}_\mu$ is an eigenvector of $\J$ associated with eigenvalue $\lambda$ satisfying
    \(
        \mu(\mu + \gamma) = \lambda.
    \)
    Consequently, the eigenvectors of $\A$ take the form
    \begin{equation}
        \pp^A_{\mu_\pm(\lambda)} = \frac{1}{\sqrt{1 + |\mu_\pm(\lambda)|^2}}
        \begin{pmatrix}
            -\mu_\pm(\lambda) \pp_\lambda \\
            \pp_\lambda
        \end{pmatrix},
    \end{equation}
    where $\pp_\lambda$ is the eigenvector of $\J$ corresponding to $\lambda$.

    Collecting these eigenvectors, we define the eigenbasis matrix $\PP_A$ such that $\A = \PP_A \M \PP_A^{-1}$
    is the spectral decomposition of $\A$.
    Because of the block structure induced by the eigenvectors and eigenvalues,
    we write the diagonal matrix $\M$ as
    \begin{equation}
        \M =
        \begin{pmatrix}
            \M_+ & \OO \\
            \OO & \M_-
        \end{pmatrix},
        \quad \text{with} \quad
        \M_\pm = \mu_\pm(\Lamb).
    \end{equation}

    The eigenbasis transformation $\PP_A$ and its inverse can be expressed in terms of $\PP$:
    \begin{equation}
        \begin{pmatrix}
            \pp^A_{\mu_+(\lambda)} & \pp^A_{\mu_-(\lambda)}
        \end{pmatrix}
        =
        \begin{pmatrix}
            \pp_\lambda & \OO \\
            \OO & \pp_\lambda
        \end{pmatrix}
        \begin{pmatrix}
            -\mu_+(\lambda) & -\mu_-(\lambda) \\
            1 & 1
        \end{pmatrix}
        \begin{pmatrix}
            (1 + |\mu_+(\lambda)|^2)^{-\frac{1}{2}} & 0 \\
            0 & (1 + |\mu_-(\lambda)|^2)^{-\frac{1}{2}}
        \end{pmatrix}.
    \end{equation}
    Extending this block structure to all eigenvalues $\lambda \in \sig(\Lamb)$ yields
    \begin{align}
        \PP_A &=
        \begin{pmatrix}
            \PP & \OO \\
            \OO & \PP
        \end{pmatrix}
        \begin{pmatrix}
            -\M_+ & -\M_- \\
            \I & \I
        \end{pmatrix}
        \begin{pmatrix}
            \N_+ & \OO \\
            \OO & \N_-
        \end{pmatrix},
        && \text{where} \quad
        \N_\pm = \left[\I + |\mu_\pm(\Lamb)|^2 \right]^{-\frac{1}{2}},
        \\
        \PP_A^{-1} &=
        \begin{pmatrix}
            \N_+^{-1} & \OO \\
            \OO & \N_-^{-1}
        \end{pmatrix}
        \begin{pmatrix}
            \Delta \M^{-1} & \OO \\
            \OO & \Delta \M^{-1}
        \end{pmatrix}
        \begin{pmatrix}
            -\I & -\M_- \\
            \I & \M_+
        \end{pmatrix}
        \begin{pmatrix}
            \PP^{-1} & \OO \\
            \OO & \PP^{-1}
        \end{pmatrix},
        && \text{with} \quad \Delta \M = \M_+ - \M_-.
    \end{align}

    Using these, the matrix exponential $e^{-\A t}$ can be expressed as
    \begin{align}
        e^{-\A t} &=
        \begin{pmatrix}
            \PP & \OO \\
            \OO & \PP
        \end{pmatrix}
        \begin{pmatrix}
            G_1(\Lamb, t) & G_2(\Lamb, t) \\
            G_3(\Lamb, t) & G_4(\Lamb, t)
        \end{pmatrix}
        \begin{pmatrix}
            \PP^{-1} & \OO \\
            \OO & \PP^{-1}
        \end{pmatrix},
        \\
        G_1(\Lamb, t) &= e^{-\frac{\gamma}{2} t} \left[\cosh\left(\frac{\Delta \M}{2} t\right) - \gamma \Delta \M^{-1} \sinh\left(\frac{\Delta \M}{2} t\right)\right],
        \\
        G_2(\Lamb, t) &= -2 e^{-\frac{\gamma}{2} t} \Lamb \Delta \M^{-1} \sinh\left(\frac{\Delta \M}{2} t\right),
        \\
        G_3(\Lamb, t) &= 2 e^{-\frac{\gamma}{2} t} \Delta \M^{-1} \sinh\left(\frac{\Delta \M}{2} t\right),
        \\
        G_4(\Lamb, t) &= e^{-\frac{\gamma}{2} t} \left[\cosh\left(\frac{\Delta \M}{2} t\right) + \gamma \Delta \M^{-1} \sinh\left(\frac{\Delta \M}{2} t\right)\right].
    \end{align}

    Finally, the dynamics of $\x$ can be written as
    \begin{equation}
        \x_t = G_3(\J, t) \vv_0 + G_4(\J, t) \x_0 + \int_0^t G_3(\J, t-s) \g_s \, ds,
    \end{equation}
    where $\x_0$ and $\vv_0$ denote the initial state and velocity.

    Assuming a damped system where the homogeneous solution vanishes over time,
    the dynamics is governed solely by the special solution (forcing).
    Setting $G := G_3$, the system evolves as
    \begin{equation}
        \label{eq:apx_kernel_gen}
        \x_t = \int_0^t G(\J, t-s) \g_s \, ds,
        \quad \text{where} \quad
        G(\lambda, t) = e^{-\frac{\gamma}{2} t} \frac{\sinh(\theta(\lambda) t)}{\theta(\lambda)},
        \quad
        \theta(\lambda) = \sqrt{\left(\frac{\gamma}{2}\right)^2 + \lambda}.
    \end{equation}

    \subsubsection{Squared Deviation with Gaussian Forcing}

    For both Gaussian and periodic forcing cases, considering a reduced two-dimensional system,  
    we utilize the explicit kernel from \eqref{eq:apx_kernel_gen} along with equations \eqref{eq:apx_var_g},  
    to compute the parameters $v_0$ \eqref{eq:apx_v0_g} and $\Delta v$ \eqref{eq:apx_K_w_g}.

    To compute $v_0$, we first evaluate the integrals
    \begin{align}
        I_\pm &= \int_0^\infty |G(\lambda_\pm, t)|^2 \, dt \\
        &= \int_0^\infty e^{-\gamma t} \left|\frac{\sinh(\theta(\lambda_\pm) t)}{\theta(\lambda_\pm)}\right|^2 dt \\
        &= \frac{1}{4 |\theta(\lambda_\pm)|^2} \int_0^\infty e^{-\gamma t} \left| e^{\theta(\lambda_\pm) t} - e^{-\theta(\lambda_\pm) t} \right|^2 dt \\
        &= \frac{1}{4 |\theta(\lambda_\pm)|^2} \int_0^\infty e^{-\gamma t} \left[ e^{2 \mathrm{Re}(\theta(\lambda_\pm)) t} + e^{-2 \mathrm{Re}(\theta(\lambda_\pm)) t} - e^{2 i \mathrm{Im}(\theta(\lambda_\pm)) t} - e^{-2 i \mathrm{Im}(\theta(\lambda_\pm)) t} \right] dt \\
        &= \frac{1}{4 |\theta(\lambda_\pm)|^2} \left[
            \frac{1}{\gamma - 2 \mathrm{Re}(\theta(\lambda_\pm))} + \frac{1}{\gamma + 2 \mathrm{Re}(\theta(\lambda_\pm))}
        - \frac{1}{\gamma - 2 i \mathrm{Im}(\theta(\lambda_\pm))} - \frac{1}{\gamma + 2 i \mathrm{Im}(\theta(\lambda_\pm))} \right] \\
        &= \frac{\gamma}{2 |\theta(\lambda_\pm)|^2} \left[
            \frac{1}{\gamma^2 - 4 \mathrm{Re}(\theta(\lambda_\pm))^2} - \frac{1}{\gamma^2 + 4 \mathrm{Im}(\theta(\lambda_\pm))^2}
        \right] \\
        &= \frac{\gamma}{2 |\theta(\lambda_\pm)|^2} \frac{4\left(\mathrm{Re}(\theta(\lambda_\pm))^2 + \mathrm{Im}(\theta(\lambda_\pm))^2\right)}{\left(\gamma^2 - 4 \mathrm{Re}(\theta(\lambda_\pm))^2\right)\left(\gamma^2 + 4 \mathrm{Im}(\theta(\lambda_\pm))^2\right)} \\
        &= \frac{2 \gamma}{\left(\gamma^2 - 4 \mathrm{Re}(\theta(\lambda_\pm))^2\right) \left(\gamma^2 + 4 \mathrm{Im}(\theta(\lambda_\pm))^2\right)} \\
        &= \frac{2 \gamma}{\gamma^4 - 4 \gamma^2 \left(\mathrm{Re}(\theta(\lambda_\pm))^2 - \mathrm{Im}(\theta(\lambda_\pm))^2 \right) - 16 \left(\mathrm{Re}(\theta(\lambda_\pm)) \mathrm{Im}(\theta(\lambda_\pm))\right)^2} \\
        &= \frac{2 \gamma}{\gamma^4 - 4 \gamma^2 \mathrm{Re}\left(\theta(\lambda_\pm)^2\right) - 4 \mathrm{Im}\left(\theta(\lambda_\pm)^2\right)^2} \\
        &= \frac{2 \gamma}{\gamma^4 - 4 \gamma^2 \left( \left(\frac{\gamma}{2}\right)^2 + \mathrm{Re}(\lambda_\pm) \right) - 4 \mathrm{Im}(\lambda_\pm)^2}
        ,\qquad \text{since } \theta(\lambda_\pm) = \sqrt{\left(\frac{\gamma}{2}\right)^2 + \lambda_\pm} \\
        \Rightarrow \quad
        I_\pm &= - \frac{\gamma}{2} \frac{1}{\gamma^2 \mathrm{Re}(\lambda_\pm) + \mathrm{Im}(\lambda_\pm)^2}.
    \end{align}

    Thus, the squared deviation for the normal system is
    \begin{equation}
        v_0 = 2 \delta \left[I_+ + I_- \right]
        = - \delta \gamma \left[
            \frac{1}{\gamma^2 \mathrm{Re}(\lambda_+) + \mathrm{Im}(\lambda_+)^2} + \frac{1}{\gamma^2 \mathrm{Re}(\lambda_-) + \mathrm{Im}(\lambda_-)^2}
        \right].
    \end{equation}

    The non-normal excess term $\Delta \nu$ \eqref{eq:apx_K_w_g} is obtained from $v_0$ and the integral
    \begin{align}
        J &= \int_0^\infty |G(\lambda_+, t) - G(\lambda_-, t)|^2 dt \\
        &= I_+ + I_- - 2 \int_0^\infty \mathrm{Re}\left( G(\lambda_+, t) G(\lambda_-, t)^* \right) dt \\
        &= I_+ + I_- - 2 \mathrm{Re}(I),
    \end{align}
    where
    \begin{align}
        I &= \int_0^\infty G(\lambda_+, t) G(\lambda_-, t)^* dt \\
        &= \frac{1}{4 \theta(\lambda_+) \theta(\lambda_-)^*} \int_0^\infty e^{-\gamma t} \left(e^{\theta(\lambda_+) t} - e^{-\theta(\lambda_+) t}\right) \left(e^{\theta(\lambda_-)^* t} - e^{-\theta(\lambda_-)^* t}\right) dt \\
        &= \frac{1}{4 \theta(\lambda_+) \theta(\lambda_-)^*} \Bigg[
            \frac{1}{\gamma - (\theta(\lambda_+) + \theta(\lambda_-)^*)} + \frac{1}{\gamma + (\theta(\lambda_+) + \theta(\lambda_-)^*)} \\
        &\quad - \frac{1}{\gamma - (\theta(\lambda_+) - \theta(\lambda_-)^*)} - \frac{1}{\gamma + (\theta(\lambda_+) - \theta(\lambda_-)^*)}
        \Bigg] \\
        &= \frac{\gamma}{2 \theta(\lambda_+) \theta(\lambda_-)^*} \left[
            \frac{1}{\gamma^2 - (\theta(\lambda_+) + \theta(\lambda_-)^*)^2} - \frac{1}{\gamma^2 - (\theta(\lambda_+) - \theta(\lambda_-)^*)^2}
        \right] \\
        &= \frac{2 \gamma}{\gamma^4 - 2 \gamma^2 \left(\theta(\lambda_+)^2 + \theta(\lambda_-)^{*2}\right) + \left(\theta(\lambda_+)^2 - \theta(\lambda_-)^{*2}\right)^2} \\
        &= - \frac{2 \gamma}{2 \gamma^2 (\lambda_+ + \lambda_-^*) - (\lambda_+ - \lambda_-^*)^2}.
    \end{align}

    Hence, the generic solution for the squared deviation in the reduced two-dimensional system under Gaussian forcing reads
    \begin{equation}
        v_\infty = v_0 \left[ 1 + K^2 \Delta v \right],
    \end{equation}
    with
    \begin{equation}
        v_0 = 2 \delta \left[I_+ + I_- \right], 
        \quad
        I_\pm = - \frac{\gamma}{2} \frac{1}{\gamma^2 \mathrm{Re}(\lambda_\pm) + \mathrm{Im}(\lambda_\pm)^2},
    \end{equation}
    and
    \begin{equation}
        \Delta v = 1 - 2 \frac{\mathrm{Re}(I)}{I_+ + I_-},
        \quad
        I = - \frac{2 \gamma}{2 \gamma^2 (\lambda_+ + \lambda_-^*) - (\lambda_+ - \lambda_-^*)^2}.
    \end{equation}

    For a given triplet $(\kappa, \lambda_+, \lambda_-)$,
    this quantifies the amplitude of the amplification mechanism induced by the non-normality of the system,
    recalling that
    \(
        K = |\kappa - \kappa^{-1}|/2.
    \)

    \subsubsection{Squared Deviation with Sinusoidal Forcing}

    When the forcing is sinusoidal rather than Gaussian, we compute the Fourier transform of the kernel \eqref{eq:apx_kernel_gen}, given by
    \begin{equation}
        \hat{G}(\lambda, \omega) = \frac{1}{\sqrt{2\pi}} \frac{1}{i \gamma \omega - \lambda - \omega^2}.
    \end{equation}
    The squared modulus for each eigenvalue is therefore
    \begin{equation}
        I_\pm := \left| \hat{G}(\lambda_\pm, \omega) \right|^2
        = \frac{1}{2\pi} \frac{1}{\left(\mathrm{Re}(\lambda_\pm) + \omega^2\right)^2 + \left(\mathrm{Im}(\lambda_\pm) + \gamma \omega \right)^2}.
    \end{equation}
    The squared difference between the two kernels reads
    \begin{equation}
        J := \left| \hat{G}(\lambda_+, \omega) - \hat{G}(\lambda_-, \omega) \right|^2 = 2 \pi I_+ I_- \, |\lambda_+ - \lambda_-|^2.
    \end{equation}

    Hence, the amplification in this generic framework is expressed as
    \begin{equation}
        v_\infty = v_0 \left[ 1 + K^2 \Delta v \right],
    \end{equation}
    where the normal squared deviation is
    \begin{equation}
        v_0 = \delta \pi \left[ I_+ + I_- \right],
        \quad \text{with} \quad
        I_\pm = \frac{1}{2\pi} \frac{1}{\left(\mathrm{Re}(\lambda_\pm) + \omega^2\right)^2 + \left(\mathrm{Im}(\lambda_\pm) + \gamma \omega\right)^2},
    \end{equation}
    and
    \begin{equation}
        \Delta v = 2 \pi |\lambda_+ - \lambda_-|^2 \frac{I_+ I_-}{I_+ + I_-}.
    \end{equation}

    As for the Gaussian forcing case, for a given triplet $(\kappa, \lambda_+, \lambda_-)$,
    one can quantify the amplitude of the amplification induced by the system's non-normality,
    recalling that
    \(
        K = \sqrt{(\kappa^2 - 1)/2}.
    \)

    \subsection{Summary: Squared Deviation}
    \label{apx:summary_deviation}

    We have shown that, without loss of generality, if a system is non-normal and subject to a time-dependent forcing,
    its squared deviation can be expressed as
    \begin{equation}
        v_\infty = v_0 \left[1 + K^2 \Delta v \right],
    \end{equation}
    where
    \begin{itemize}
        \item $v_0$ is the squared deviation in the normal case, depending only on the spectrum of the system matrix;
        \item $K$ is a degree of non-normality, which satisfies $K = 0$ when the system is normal and increases monotonically with the condition number of the eigenbasis transformation matrix of the linear system;
        \item $\Delta v$ is a coefficient that depends solely on the spectrum of the matrix.
    \end{itemize}

    We distinguish two forcing scenarios:

    \begin{itemize}
        \item \textbf{Gaussian forcing:}
        \begin{align}
            v_0 &= 2 \delta \left[ I_+ + I_- \right], \quad
            && I_\pm = -\frac{\gamma}{2} \frac{1}{\gamma^2 \mathrm{Re}(\lambda_\pm) + \mathrm{Im}(\lambda_\pm)^2},
            \label{eq:apx_v0_gauss} \\
            \Delta v &= 1 - 2 \frac{\mathrm{Re}(I)}{I_+ + I_-}, \quad
            && I = -\frac{2 \gamma}{2 \gamma^2 (\lambda_+ + \lambda_-^*) - (\lambda_+ - \lambda_-^*)^2}, \\
            K &= \frac{1}{2} \left| \kappa - \kappa^{-1} \right|.
            \label{eq:apx_dv_gauss}
        \end{align}

        \item \textbf{Sinusoidal forcing:}
        \begin{align}
            v_0 &= \delta \pi \left[ I_+ + I_- \right], \quad
            && I_\pm = \frac{1}{2 \pi} \frac{1}{\left(\mathrm{Re}(\lambda_\pm) + \omega^2\right)^2 + \left(\mathrm{Im}(\lambda_\pm) + \gamma \omega\right)^2},
            \label{eq:apx_v0_sin} \\
            \Delta v &= 2 \pi |\lambda_+ - \lambda_-|^2 \frac{I_+ I_-}{I_+ + I_-}, \\
            K &= \sqrt{\frac{\kappa^2 - 1}{2}}.
            \label{eq:apx_dv_sin}
        \end{align}
    \end{itemize}

    Thus, the impact of non-normality and the associated amplification mechanism
    can be completely characterized by the triplet $(\kappa, \lambda_+, \lambda_-)$.

    While a full three-dimensional phase diagram capturing regimes of low, moderate, and critical non-normality is challenging to visualize,
    we propose to study these results in the reduced $(\alpha, \beta)$ parameter space,
    which enables constructing clear phase diagrams in $(K, \alpha)$ and $(K, \beta)$,
    providing intuitive insight into the system's behavior under varying degrees of non-normality.

    \section{Unifying Amplification Framework}

    In this section, we use the results derived in \ref{apx:summary_deviation} to develop a unifying structure within the $\alpha/\beta$-framework.  
    This framework provides explicit closed-form expressions that allow us to identify the origin of amplification in the system,
    i.e., whether it arises from criticality, resonance, or non-normality.

    First, we analyse overdamped systems driven by a Gaussian white noise, which offers a simplified
    set-up  for estimating how strongly a system must be non-normal, relative to its proximity to criticality, to exhibit non-normal amplification.

    Next, we derive the squared deviations for both Gaussian and sinusoidal forcing in the $\alpha$ and $\beta$ reduced forms.

    Finally, we summarize the results in a table, enabling a straightforward assessment of whether a system's amplification is primarily due to non-normality or rather driven solely by criticality and resonance mechanisms.

    \subsection{Overdamped System Driven by a Gaussian White Noise}

    In the overdamped regime, the kernel simplifies to an exponential form, i.e.,
    \begin{equation}
        G(\lambda, t) = e^{\lambda t}.
    \end{equation}
    Thus, the square of the Frobenius norm of the kernel \eqref{eq:apx_2d_frob} is given by
    \begin{equation}
        \|G(\J, t)\|_F^2 = |e^{\lambda_+ t}|^2 + |e^{\lambda_- t}|^2 + K^2 \left| e^{\lambda_+ t} - e^{\lambda_- t} \right|^2,
    \end{equation}
    which defines the noise kernel for each Gaussian innovation.

    In the normal case ($K=0$), the kernel decreases monotonically over time.
    However, when the system is non-normal ($K > 0$),
    the interplay between the exponential terms,
    i.e., the difference $e^{\lambda_+ t} - e^{\lambda_- t}$,
    can produce transient growth of each noise innovation,
    provided that the eigenvalues are distinct (i.e., $\lambda_+ \neq \lambda_-$).

    Our goal here is to determine the condition under which the Frobenius norm ceases to be monotonically decreasing,
    that is, when a local extremum appears.
    To this end, we analyze the problem in the two $\alpha/\beta$-frameworks,
    deriving the corresponding conditions in the two-dimensional parameter spaces $(\alpha, K)$ and $(\beta, K)$.

    \subsubsection{$\alpha$-Form}

    For the $\alpha$-form, the eigenvalues are given by $\lambda_\pm = -\alpha \pm \chi$,
    where $\chi = 1$ or $i$. We analyze the extremum of $\|G(\J,t)\|_F^2$ for each case.

    \begin{itemize}
        \item \textbf{$\chi=1$} -- Here, we write
        \begin{equation}
            \|G(\J,t)\|_F^2 = \|e^{\Gam_1^\alpha t}\|_F^2 = 2(1+K^2) e^{-2\alpha t} \left[\cosh(2t) - z\right],
        \end{equation}
        which yields the time derivative
        \begin{equation}
            \frac{d}{dt} \|G(\J,t)\|_F^2 = -4(1+K^2) e^{-2\alpha t} \left[ \alpha \cosh(2t) - \sinh(2t) - 2\alpha z \right],
        \end{equation}
        where we defined
        \begin{equation}
            z = \frac{K^2}{K^2 + 1} \quad \Rightarrow \quad K = \sqrt{\frac{z}{1-z}}.
        \end{equation}

        Defining $T = \tanh(t)$, recall the identities
        \begin{equation}
            \cosh(2t) = \frac{1 + T^2}{1 - T^2}, \quad \sinh(2t) = \frac{2T}{1 - T^2},
        \end{equation}
        so setting the derivative to zero gives the quadratic equation
        \begin{equation}
            \alpha(1+z) T^2 - 2T + \alpha(1 - z) = 0,
        \end{equation}
        with solutions
        \begin{equation}
            T_\pm = \frac{1}{\alpha(1+z)} \left[ 1 \pm \sqrt{1 - \alpha^2 (1 - z^2)} \right].
        \end{equation}

        Thus, real solutions exist (and the kernel is not monotonically decreasing) if
        \begin{equation}
            1 - \alpha^2 (1 - z^2) \geq 0 \quad \Rightarrow \quad z \geq z_c = \frac{\sqrt{\alpha^2 - 1}}{\alpha}.
        \end{equation}
        Equivalently, the critical degree of non-normality is
        \begin{equation}
            K \geq K_c = \sqrt{\frac{\sqrt{\alpha^2 - 1}}{\alpha - \sqrt{\alpha^2 - 1}}} =
            \begin{cases}
                \sqrt{2}(\alpha - 1) + \mathcal{O}((\alpha - 1)^2), & \alpha - 1 \gg 0, \\
                (2(\alpha - 1))^{1/4} + \mathcal{O}((\alpha - 1)^{3/4}), & 0 < \alpha - 1 \ll 1.
            \end{cases}
        \end{equation}

        \item \textbf{$\chi = i$} -- In this case,
        \begin{equation}
            \|G(\J,t)\|_F^2 = 2(1+K^2) e^{-2\alpha t} \left[ 1 - z \cos(2t) \right],
        \end{equation}
        and
        \begin{equation}
            \frac{d}{dt} \|G(\J,t)\|_F^2 = -4(1+K^2) e^{-2\alpha t} \left[ \alpha - \alpha z \cos(2t) - z \sin(2t) \right].
        \end{equation}

        Using $T = \tan(t)$ and the identities
        \begin{equation}
            \cos(2t) = \frac{1 - T^2}{1 + T^2}, \quad \sin(2t) = \frac{2T}{1 + T^2},
        \end{equation}
        the condition for extrema takes the form of the quadratic equation
        \begin{equation}
            \alpha(1+z) T^2 - 2 z T + \alpha(1 - z) = 0,
        \end{equation}
        with solutions
        \begin{equation}
            T_\pm = \frac{z}{\alpha(1+z)} \left[ 1 \pm \sqrt{1 - \alpha^2 \frac{1 - z^2}{z^2}} \right].
        \end{equation}

        Real solutions exist if
        \begin{equation}
            1 - \alpha^2 \frac{1 - z^2}{z^2} \geq 0 \quad \Rightarrow \quad z \geq z_c = \frac{\alpha}{\sqrt{\alpha^2 + 1}},
        \end{equation}
        giving the critical non-normality
        \begin{equation}
            K \geq K_c = \sqrt{\frac{\alpha}{\sqrt{\alpha^2 + 1} - \alpha}} =
            \begin{cases}
                \sqrt{2}\alpha + \mathcal{O}(\alpha^{-5}), & \alpha \gg 1, \\
                \sqrt{\alpha} + \mathcal{O}(\alpha^{5/4}), & 0 < \alpha \ll 1.
            \end{cases}
        \end{equation}
    \end{itemize}

    In both cases, if the system is far from criticality (i.e., $\alpha \gg 1$),
    transient deviation in noise increments appear when $K > K_c \sim \alpha$.
    Near criticality,
    the system exhibits strong non-normal effects when 
    \begin{equation}
        K > K_c \sim (\alpha - \alpha_c)^m,
    \end{equation}
    where the critical value $\alpha_c=1$ for $\chi=1$, and $\alpha_c=0$ for $\chi=i$,
    and the exponent $m=1/4$ for $\chi=1$ and $m=1/2$ for $\chi=i$.
    
    \subsubsection{$\beta$-form}

    For the $\beta$-form, the eigenvalues are given by $\lambda_\pm = -1 \pm \beta \chi$.
    We analyze the extremum of $\|G(\J,t)\|_F^2$ for the two cases $\chi=1$ and $\chi=i$.

    \begin{itemize}
        \item \textbf{$\chi=1$} -- In this case, we have
        \begin{equation}
            \|G(\J,t)\|_F^2 = \|e^{\Gam_1^\beta t}\|_F^2 = 2(1+K^2) e^{-2 t} \left[\cosh(2 \beta t) - z\right],
        \end{equation}
        and its time derivative is
        \begin{equation}
            \frac{d}{dt} \|G(\J,t)\|_F^2 = -4(1+K^2) e^{-2 t} \left[\cosh(2 \beta t) - \beta \sinh(2 \beta t) - z \right].
        \end{equation}

        Defining \(T = \tanh(\beta t)\), we use
        \begin{equation}
            \cosh(2 \beta t) = \frac{1 + T^2}{1 - T^2}, \quad \sinh(2 \beta t) = \frac{2 T}{1 - T^2},
        \end{equation}
        so the extrema satisfy the quadratic equation
        \begin{equation}
            (1 + z) T^2 - 2 \beta T + 1 - z = 0,
        \end{equation}
        with solutions
        \begin{equation}
            T_\pm = \frac{1}{1 + z} \left[ \beta \pm \sqrt{\beta^2 - (1 - z^2)} \right].
        \end{equation}

        Real solutions exist if
        \begin{equation}
            z \geq z_c = \sqrt{1 - \beta^2} \quad \Rightarrow \quad
            K \geq K_c = \sqrt{\frac{\sqrt{1 - \beta^2}}{1 - \sqrt{1 - \beta^2}}} =
            \begin{cases}
                \frac{\sqrt{2}}{\beta} + \mathcal{O}(1), & 0 < \beta \ll 1, \\
                (1 - \beta)^{1/4} + \mathcal{O}((1 - \beta)^{3/4}), & 0 < 1 - \beta \ll 1.
            \end{cases}
        \end{equation}

        \item \textbf{$\chi=i$} -- Here,
        \begin{equation}
            \|G(\J,t)\|_F^2 = 2(1+K^2) e^{-2 t} \left[1 - z \cos(2 \beta t)\right],
        \end{equation}
        and
        \begin{equation}
            \frac{d}{dt} \|G(\J,t)\|_F^2 = -4(1+K^2) e^{-2 t} \left[1 - z \cos(2 \beta t) - z \beta \sin(2 \beta t)\right].
        \end{equation}

        Using \(T = \tan(\beta t)\) and the identities
        \begin{equation}
            \cos(2 \beta t) = \frac{1 - T^2}{1 + T^2}, \quad \sin(2 \beta t) = \frac{2 T}{1 + T^2},
        \end{equation}
        the extrema satisfy
        \begin{equation}
            (1 + z) T^2 - 2 z \beta T + 1 - z = 0,
        \end{equation}
        with solutions
        \begin{equation}
            T_\pm = \frac{1}{1 + z} \left[z \beta \pm \sqrt{(z \beta)^2 - (1 - z^2)}\right].
        \end{equation}

        Real solutions exist if
        \begin{equation}
            z \geq z_c = \frac{1}{\sqrt{1 + \beta^2}} \quad \Rightarrow \quad
            K \geq K_c = \sqrt{\frac{1}{\sqrt{1 + \beta^2} - 1}} =
            \begin{cases}
                \frac{\sqrt{2}}{\beta} + \mathcal{O}(\beta), & 0 < \beta \ll 1, \\
                \beta^{-1/2} + \mathcal{O}(\beta^{-3/2}), & \beta \gg 1.
            \end{cases}
        \end{equation}
    \end{itemize}

    In both cases, when the system is close to degeneracy, i.e. \(0 < \beta \ll 1\),
    the critical degree of non-normality diverges as \(K_c \sim 1/\beta\),
    so pseudo-critical behavior is observed only if \(K > K_c \sim 1/\beta\).

    For real eigenvalues (\(\chi = 1\)) near the critical value \(\beta \to 1^-\)
    (with stability requiring \(0 < \beta < 1\)),
    the critical degree of non-normality follows a power law
    \(
        K_c \sim (1 - \beta)^{1/4}.
    \)

    When the eigenvalues are complex conjugates (\(\chi = i\)),
    there is no critical point in the overdamped framework, so \(\beta\) can vary freely.
    For large \(\beta \gg 1\),
    the critical degree scales as
    \(
        K_c \sim 1/\beta^{1/2}.
    \)

    \subsubsection{Summary: Critical Degree of Non-Normality \(K_c\)}

    Considering only the overdamped scenarios, we identify a critical degree of non-normality \(K_c\) such that, for \(K > K_c\),  
    the kernel no longer decays monotonically and exhibits transient deviations.  
    These transient deviations are the source of the amplification mechanism induced by non-normality.

    For the \(\alpha\)- and \(\beta\)-forms, the critical values \(K_c\) are given by

    \begin{align}
        &\textbf{\(\alpha\)-form} \quad
        K_c = 
        \begin{cases}
            \sqrt{\frac{\sqrt{\alpha^2 - 1}}{\alpha - \sqrt{\alpha^2 - 1}}}, & \alpha > 1, \quad \chi = 1 \\[6pt]
            \sqrt{\frac{\alpha}{\sqrt{\alpha^2 + 1} - \alpha}}, & \alpha > 0, \quad \chi = i
        \end{cases}, \\
        &\textbf{\(\beta\)-form} \quad
        K_c =
        \begin{cases}
            \sqrt{\frac{\sqrt{1 - \beta^2}}{1 - \sqrt{1 - \beta^2}}}, & 0 < \beta < 1, \quad \chi = 1 \\[6pt]
            \sqrt{\frac{1}{\sqrt{1 + \beta^2} - 1}}, & \beta > 0, \quad \chi = i
        \end{cases}.
    \end{align}

    The main observations in each framework are:

    \begin{itemize}
        \item \textbf{\(\alpha\)-form} — When the system is close to criticality,  
        the critical degree of non-normality \(K_c\) tends to zero,  
        meaning that before becoming truly critical, the system exhibits pseudo-critical behavior.  
        Conversely, far from criticality, a useful heuristic is that the degree of non-normality \(K\) should exceed the distance from criticality for 
        significant transient amplification to occur.

        \item \textbf{\(\beta\)-form} — As the system approaches degeneracy (i.e., \(\beta \to 0\)),  
        the critical degree of non-normality diverges,  
        since the matrix \(\J\) tends to a scalar multiple of the identity and becomes invariant under basis transformations,  
        thus diminishing the effect of non-normality.  
        In contrast, as \(\beta\) moves away from degeneracy, \(K_c\) tends to zero,  
        implying that the system will generally exhibit non-normal amplification.
    \end{itemize}

    In conclusion, within the overdamped framework,  
    the closer a system is to criticality, the easier it is to exhibit pseudo-critical transient amplification.  
    Conversely, near degeneracy of eigenvalues, the system becomes less susceptible to non-normal transient effects.

    \subsection{Non-Normal Gain}

    We have shown that the squared deviation of a non-normal system can be expressed as
    \begin{equation}
        v_\infty = v_0 \left[1 + G\right],
    \end{equation}
    where \(v_0\) is the squared deviation in the corresponding normal system,  
    and \(G\) is the \emph{non-normal gain}, a function depending on the degree of non-normality \(K\), the eigenvalues \(\lambda_+\) and \(\lambda_-\),  
    and other system parameters such as the forcing frequency \(\omega\) (in the sinusoidal forcing case) and the damping coefficient \(\gamma\).

    By construction, \(G\) satisfies
    \(
        G(K=0, \lambda_+, \lambda_-) = 0,
    \)
    since when the system is normal (\(K=0\)), there is no non-normal amplification. For non-normal systems (\(K > 0\)), we have
    \(
        G(K, \lambda_+, \lambda_-) > 0.
    \)

    Furthermore, the non-normal gain can be decomposed as
    \begin{equation}
        G := K^2 \Delta v,
    \end{equation}
    where \(\Delta v\) is independent of the non-normality degree \(K\),  
    depending only on the eigenvalues of the system and parameters such as \(\gamma\) and \(\omega\) (for sinusoidal forcing).

    In the following, we will provide explicit expressions for \(\Delta v\) in both forcing scenarios (Gaussian and sinusoidal)  
    and for the two \(\alpha/\beta\)-forms.

    \subsubsection{$\alpha$-form}

    In the $\alpha$-form, the eigenvalues are given by $\lambda_\pm = -\alpha \pm \chi$,  
    where $\chi = 1$ or $i$. The squared deviation in the normal case,
    $v_0$ \eqref{eq:apx_v0_gauss}, and the coefficient $\Delta v$ \eqref{eq:apx_dv_gauss} can be obtained as follows for Gaussian forcing
    \begin{equation}
        v_0 = 
        \begin{cases}
            \frac{2\delta}{\gamma}\frac{\alpha}{\alpha^2 - 1},\quad \alpha > 1 & \chi = 1, \\[6pt]
            \frac{2\delta \gamma}{\gamma^2 \alpha - 1},\quad \alpha > \frac{1}{\gamma^2} & \chi = i,
        \end{cases}
        \quad\text{and}\quad
        \Delta v = 
        \begin{cases}
            \frac{\gamma^2 + \alpha}{\alpha(\gamma^2 \alpha + 1)},\quad \alpha > 1 & \chi = 1, \\[6pt]
            \frac{\gamma^2 + \alpha}{\gamma^2 (\alpha^2 + 1)},\quad \alpha > \frac{1}{\gamma^2} & \chi = i.
        \end{cases}
    \end{equation}

    In both cases, the critical point corresponds to the singularity of the normal squared deviation \(v_0\),  
    which is at
    \begin{equation}
        \alpha_c = 
        \begin{cases}
            1, & \chi=1 \\
            \gamma^{-2}, &\chi=i
        \end{cases}.
    \end{equation}
    Similarly, \(\Delta v\) reaches its maximum of $1$ at these critical points, i.e., \(\Delta v = 1\) when \(\alpha = \alpha_c\).

    Thus, the non-normal amplification mechanism is maximal at criticality.  
    When the system is far from criticality (\(\alpha \gg \alpha_c\)),  
    a useful rule of thumb to observe a non-normal gain of order one or larger (\(G \gtrsim 1\))  
    is that the degree of non-normality satisfies
    \[
    K \gtrsim (\alpha - \alpha_c),
    \]
    consistent with the estimation derived previously for the overdamped Gaussian forcing scenario.
    \newline

    For sinusoidal forcing with pulsation \(\omega\), the expressions \eqref{eq:apx_v0_sin} and \eqref{eq:apx_dv_sin} become
    \begin{equation}
        v_0 = 
        \begin{cases}
            \frac{(\omega - \alpha)^2 + 1 + (\gamma \omega)^2}{\big[(\gamma \omega)^2 - 1\big]^2 + 2\big[(\gamma \omega)^2 + 1\big](\omega^2 - \alpha)^2 + (\omega^2 - \alpha)^4}, & \chi = 1, \\[12pt]
            \frac{(\omega - \alpha)^2 + 1 + (\gamma \omega)^2}{(\gamma \omega)^4 + 2(\gamma \omega)^2\big[(\omega^2 - \alpha)^2 + 1\big] + \big[(\omega^2 - \alpha)^2 - 1\big]^2}, & \chi = i,
        \end{cases}
        \quad\text{and}\quad
        \Delta v = \frac{2}{(\omega^2 - \alpha)^2 + 1 + (\gamma \omega)^2}, \quad \chi = 1,i.
    \end{equation}

    The resonance (critical) points occur at
    \begin{equation}
        \begin{cases}
            \gamma \omega = 0, \quad \alpha_c = \omega^2 \pm 1, & \chi = 1, \\
            \gamma \omega = \pm 1, \quad \alpha_c = \omega^2, & \chi = i,
        \end{cases}
    \end{equation}
    where \(\Delta v = 1\).

    Away from resonance, \(\Delta v\) decreases quadratically with the distance from these points, similar to the Gaussian forcing case.  
    Hence, the same rule of thumb applies: to observe non-normal amplification of order one or more (\(G \gtrsim 1\)),  
    the degree of non-normality \(K\) should be at least on the order of the distance to the resonance point.

    \subsubsection{$\beta$-form}

    In the $\beta$-form, the eigenvalues are given by $\lambda_\pm = -1 \pm \beta \chi$, where $\chi = 1, i$.  
    For Gaussian forcing, the normal squared deviation $v_0$ \eqref{eq:apx_v0_gauss} and the coefficient $\Delta v$ \eqref{eq:apx_dv_gauss} are
    \begin{equation}
        v_0 = 
        \begin{cases}
            \frac{2\delta}{\gamma} \frac{1}{1 - \beta^2}, \quad 1 > \beta > 0 & \chi = 1, \\[6pt]
            \frac{2\delta \gamma}{\gamma^2 - \beta^2}, \quad \gamma > \beta > 0 & \chi = i,
        \end{cases}
        \quad\text{and}\quad
        \Delta v = 
        \begin{cases}
            \frac{\beta^2 (1 + \gamma^2)}{\gamma^2 + \beta^2}, \quad 1 > \beta > 0 & \chi = 1, \\[6pt]
            \frac{\beta^2}{\gamma^2} \frac{1 + \gamma^2}{1 + \beta^2}, \quad \gamma > \beta > 0 & \chi = i.
        \end{cases}
    \end{equation}

    In both cases, the system is degenerate at $\beta = 0$ and has a finite critical point $\beta_c$, with  
    \begin{equation}
        \beta_c = 
        \begin{cases}
        1, & \chi = 1, \\
        \gamma, & \chi = i.
        \end{cases}
    \end{equation}
    The coefficient $\Delta v$ reaches its maximum of $1$ at the critical point, i.e. $\Delta v = 1$ when $\beta = \beta_c$.  
    Thus, the non-normal amplification is maximal at criticality, as also seen in the $\alpha$-form.

    When the system is degenerate ($\beta = 0$), the coefficient $\Delta v$ is zero, implying no non-normal gain ($G=0$).  
    This degeneracy limit corresponds to the asymptotic limit $\alpha \to \infty$ in the $\alpha$-form.

    Hence, similar to the $\alpha$-form, a heuristic for observing non-normal gain of order one or more ($G \gtrsim 1$) is
    \(
        K \gtrsim 1/(\beta - \beta_c).
    \)
    \newline

    For sinusoidal forcing with pulsation $\omega$, from \eqref{eq:apx_v0_sin} and \eqref{eq:apx_dv_sin} we have
    \begin{equation}
        v_0 = 
        \begin{cases}
            \delta \frac{(\omega^2 - 1)^2 + \beta^2 + (\gamma \omega)^2}{(\gamma \omega)^4 + 2 (\gamma \omega)^2 ((\omega^2 - 1)^2 + \beta^2) + ((\omega^2 - 1)^2 - \beta^2)^2}, & \chi = 1, \\[12pt]
            \delta \frac{(\omega - 1)^2 + \beta^2 + (\gamma \omega)^2}{((\gamma \omega)^2 - \beta^2)^2 + 2 ((\gamma \omega)^2 + \beta^2)(\omega^2 - 1)^2 + (\omega^2 - 1)^4}, & \chi = i,
        \end{cases}
        \quad\text{and}\quad
        \Delta v = \frac{2 \beta^2}{(\omega^2 - 1)^2 + \beta^2 + (\gamma \omega)^2}, \quad \chi = 1,i.
    \end{equation}

    The resonance (critical) points occur at
    \begin{equation}
        \begin{cases}
            \gamma \omega = 0, \quad \beta = |\omega^2 - 1|, & \chi = 1, \\
            \beta = \gamma \omega, \quad \omega = 1 \implies \beta = \gamma, & \chi = i,
        \end{cases}
    \end{equation}
    where $\Delta v = 1$. When $\beta = 0$ (degenerate case), $\Delta v = 0$, its minimum value.  
    In the asymptotic limit $\beta \to \infty$, $\Delta v \to 2$, which is its maximal value.

    Therefore, as in the Gaussian forcing case, a useful heuristic to observe non-normal amplification of order one or larger ($G \gtrsim 1$) is
    \(
    K \gtrsim 1/\beta \; \text{for } \beta \gtrsim 0.
    \)

    \subsubsection{Summary: Non-Normal Gain \& Criticality}
    \label{apx:summary_uni}

    We have shown that the non-normal gain $G := K^2 \Delta v$ depends not only on the degree of non-normality $K$, but also on the distance to criticality (or resonance) and degeneracy.  
    In this section, we summarize the main results assuming the damping coefficient $\gamma$ is fixed, except for the sinusoidal forcing case where a critical damping $\gamma_c$ may exist.

    Table~\ref{tab:apx_summary-non-normal-gain} summarizes these findings, including the expressions for $v_0$, $\Delta v$, and critical points in the $\alpha$ and $\beta$ frameworks for Gaussian and sinusoidal forcing.

    We also recall that the degree of non-normality $K$ depends on the condition number $\kappa$ of the eigenbasis transformation matrix of $\mathbf{J}$, differing between forcing types as
    \begin{equation}
        K = 
        \begin{cases}
            \frac{1}{2} \left| \kappa - \kappa^{-1} \right|, & \text{Gaussian Forcing}, \\[6pt]
            \sqrt{\frac{\kappa^2 - 1}{2}}, & \text{Sinusoidal Forcing}.
        \end{cases}
    \end{equation}

    \begin{table}[h!]
        \centering
            \begin{tabular}{|c|c|c|c|c|c|}
            \hline

            \textbf{Form} & \textbf{Forcing Type} & \textbf{Eig. Type} & $\mathbf{\Delta v}$ & \textbf{Critical Point} & \textbf{Stab. Domain} \\
            \hline
            \multirow{4}{*}{$\alpha$-form}
            & \multirow{2}{*}{Gaussian Forcing} & $\chi = 1$ & $\frac{\gamma^2 + \alpha}{\alpha(\gamma^2\alpha + 1)}$ & $\alpha_c=1$ & $\alpha> 1$ \\
            \cline{3-6}
                                        &  & $\chi = i$ & $\frac{\gamma^2 + \alpha}{\gamma^2(\alpha^2 + 1)}$ & $\alpha_c=1/\gamma^2$ & $\alpha>1/\gamma^2$   \\
            \cline{2-6}
            & \multirow{2}{*}{Sinusoidal Forcing} & $\chi = 1$ & \multirow{2}{*}{$\frac{2}{(\omega^2-\alpha)^2 + 1 + (\gamma\omega)^2}$}  & $\gamma_c=0,\; \alpha_c = \omega^2\pm 1$ & \multirow{2}{*}{$\alpha\in \mathbb{R}\backslash\{\alpha_c\}$}  \\
            \cline{3-3}\cline{5-5}
                                              &  & $\chi = i$ &      & $\gamma_c = 1/\omega,\, \alpha_c=\omega^2$ &   \\
            \hline
            \multirow{4}{*}{$\beta$-form}
            & \multirow{2}{*}{Gaussian Forcing} & $\chi = 1$ & $\frac{\beta^2(1+\gamma^2)}{\gamma^2 + \beta^2}$ & $\beta_c=1$ & $\beta\in (1,0]$  \\
            \cline{3-6}
                                        &    & $\chi = i$ & $\frac{\beta^2}{\gamma^2}\frac{1+\gamma^2}{1+\beta^2}$ & $\beta_c = \gamma$ & $\beta\in (\gamma,0]$  \\
            \cline{2-6}
            & \multirow{2}{*}{Sinusoidal Forcing} & $\chi = 1$ & \multirow{2}{*}{$\frac{2\beta^2}{(\omega^2 - 1)^2 + \beta^2 + (\gamma\omega)^2}$}  & $\gamma_c=0,\, \beta_c=|\omega^2-1|$ & \multirow{2}{*}{$\beta\in \mathbb{R}\backslash\{\beta_c\}$} \\
            \cline{3-3}\cline{5-5}
                                              &  & $\chi = i$ &      & $\omega=1\, \beta_c = \gamma$ &  \\
            \hline
            \end{tabular}
        \caption{Summary Table of the Non-Normal Gain and Critical Points. \label{tab:apx_summary-non-normal-gain}}
    \end{table}

    \section{Applications to Existing Models}

    The goal of this section is to demonstrate how the unifying framework summarized in \ref{apx:summary_uni} can be applied to existing theoretical models.
    This approach allows researchers studying critical systems and bifurcations to estimate whether their systems exhibit non-normality,
    and whether non-normal amplification mechanisms dominate over classical criticality or resonance effects.

    Such insight will enhance the understanding of amplification phenomena in complex natural systems by clarifying the underlying mechanisms.

    We will first apply the framework to earthquake triggering by analyzing a reduced model for frictional amorphous solids.
    This example will show that the system is indeed non-normal and we will identify regions in parameter space where non-normal amplification dominates,
    beyond the usual criticality and resonance scenarios.

    Next, we will discuss a system from quantum optics known for its non-Hermitian nature
    -- a necessary condition for non-normality -- 
    and exhibiting Parity-Time (PT) symmetry breaking, which corresponds to a Hopf bifurcation.
    This example highlights the interplay of non-normality and criticality in such physical systems.

    \subsection{Frictional Amorphous Solids}

    The mechanism behind the giant amplification of small perturbations in frictional amorphous solids 
    has been suggested as crucial for understanding catastrophic events such as remote earthquake triggering \cite{Charan2020}.  
    Focusing on systems whose dynamics are non-Hamiltonian,  
    this model explores the influence of non-potential forces and dissipation on stability and sensitivity to external perturbations.  
    In particular, it examines the behavior of eigenvalues and eigenvectors near critical points,  
    where small parameter changes can induce large system responses.

    We first recall the reduced form of the model proposed in \cite{Charan2020},  
    then derive its non-normal structure, demonstrating that it naturally fits within the $\beta$-framework.  
    Finally, we show how focusing solely on criticality in previous studies has overlooked significant portions of the phase diagram where non-normal amplification dominates,  
    revealing a richer understanding of the system’s amplification mechanisms.

    \subsubsection{Reduced Form}

    To analyze the sensitivity of the system to external perturbations near criticality,  
    a reduced normal form was proposed in \cite{Charan2020, Charan2021Remote}.  
    For a two-dimensional system, the dynamics can be described by a set of linear differential equations involving two independent parameters.  
    The general form is given by
    \begin{equation}
        \label{eq:normal_form_2}
        \ddot{\x} + \gamma \dot{\x} = \J \x + \g,
        \quad \text{where} \quad
        \J =
        \begin{pmatrix}
            -1 + \delta & -\eta \\
            \eta & -1 - \delta
        \end{pmatrix},
        \quad \delta, \eta > 0,
    \end{equation}
    with $\x = (x, y)$ representing the system state vector, and $\g$ a time-dependent forcing term.

    \subsubsection{Non-Normal Decomposition}
    \label{apx:non_normal_decomp}

    This equation represents the system's behavior via the Jacobian matrix, decomposed into symmetric and skew-symmetric parts.  
    The eigenvalues ($\lambda_\pm$) and eigenvectors ($\pp_\pm$) are given by
    \begin{align}
        &\lambda_\pm = -1 \pm \sqrt{\delta^2 - \eta^2}, \\
        &\pp_+ = \frac{1}{\sqrt{1 + |\nu|^2}}
        \begin{pmatrix}
            1 \\
            |\nu| e^{i \phi}
        \end{pmatrix}, \quad
        \pp_- = \frac{1}{\sqrt{1 + |\nu|^2}}
        \begin{pmatrix}
            |\nu| \\
            e^{-i \phi}
        \end{pmatrix}, \\
        &\text{where} \quad
        \nu = \frac{\delta}{\eta} + \sqrt{\left(\frac{\delta}{\eta}\right)^2 - 1},
        \quad
        \phi = \arg(\nu).
    \end{align}
    These eigenvalues and eigenvectors characterize the system's stability and response to perturbations.

    Assuming the initial conditions $\x(0) = 0$ and $\dot{\x}(0) = 0$, the solution to \eqref{eq:normal_form_2} reads
    \begin{equation}
        \label{eq:gen_sol}
        \x_t = \int_0^t G(\J, t-s)\,\g_s \, ds,
        \quad \text{where} \quad
        G(\lambda, t) = e^{-\gamma t / 2} \frac{\sinh(\theta t)}{\theta},
        \quad \theta := \sqrt{\left(\frac{\gamma}{2}\right)^2 - \lambda}.
    \end{equation}

    To analyze non-normality, we express the condition number $\kappa$ of the eigenbasis transformation $\PP = (\pp_+, \pp_-)$ in terms of the control parameters $\delta$ and $\eta$.  
    Performing the singular value decomposition (SVD) of $\PP = \U \Sig \V^\dag$ yields
    \begin{equation}
        \Sig = \frac{|1 + \nu|}{\sqrt{1 + |\nu|^2}}
        \begin{pmatrix}
            1 & 0 \\
            0 & \kappa^{-1}
        \end{pmatrix},
        \quad
        \U = \frac{1}{\sqrt{2}}
        \begin{pmatrix}
            1 & 1 \\
            1 & -1
        \end{pmatrix},
        \quad
        \V = \frac{1}{\sqrt{2}}
        \begin{pmatrix}
            1 & 1 \\
            e^{i \phi} & -e^{i \phi}
        \end{pmatrix},
        \quad
        \kappa = \left| \frac{\nu + 1}{\nu - 1} \right|.
    \end{equation}

    Thus, we can write
    \begin{equation}
        \PP = \sigma_+ \left[ \U \V^\dag + (\kappa^{-1} - 1) \uu_- \vv_-^\dag \right],
    \end{equation}
    where $\sigma_+$ is the largest singular value of $\PP$, and $\uu_-$ and $\vv_-$ are the second columns of $\U$ and $\V$, respectively.

    Applying the unitary rotation $\V \U^\dag$ gives
    \begin{equation}
        \V \U^\dag \PP = \sigma_+ \left[ \I + (\kappa^{-1} - 1) \vv_- \vv_-^\dag \right],
    \end{equation}
    matching the setup used throughout the present manuscript and its appendices.

    The non-normal mode is then identified as $\nn = \vv_-$, with the corresponding reactive mode orthogonal to it, $\rr = \vv_+$.  
    Hence, the eigenbasis transformation can be expressed as
    \begin{equation}
        \PP = \sigma_+ \left( \rr \rr^\dag + \kappa^{-1} \nn \nn^\dag \right).
    \end{equation}

    We summarize the control parameters in the $\beta$-form, linking $(\delta, \eta)$ to $(\beta, \kappa)$ as
    \begin{equation}
        \beta = \sqrt{|\delta^2 - \eta^2|}, \quad
        \kappa = \sqrt{\left| \frac{\delta + \eta}{\delta - \eta} \right|}.
    \end{equation}
    When $\delta > \eta$, the system is in the $\chi=1$ case (real eigenvalues),  
    and for $\delta < \eta$, the system is in the $\chi = i$ case (complex conjugate eigenvalues).

    As $\delta \to \eta^+$, the system approaches a Hamiltonian Hopf bifurcation and exhibits strong non-normality, with $\kappa \to \infty$.  
    Since both $\delta$ and $\eta$ are free parameters, one can independently control $\beta$ and $\kappa$ via
    \begin{equation}
        \begin{cases}
            \delta = \frac{\beta}{2} \left( \kappa + \kappa^{-1} \right), \\
            \eta = \frac{\beta}{2} \left( \kappa - \kappa^{-1} \right),
        \end{cases}
        \quad \text{if} \quad \delta > \eta,
        \quad \text{and} \quad
        \begin{cases}
            \delta = \frac{\beta}{2} \left( \kappa - \kappa^{-1} \right), \\
            \eta = \frac{\beta}{2} \left( \kappa + \kappa^{-1} \right),
        \end{cases}
        \quad \text{if} \quad \eta > \delta.
    \end{equation}

    \subsubsection{Misleading Non-Normality}

   Reef.  \cite{Charan2020} defined the parameter
    \begin{equation}
        \epsilon = 1 - \frac{\eta}{\delta}, \quad \delta > \eta,
    \end{equation}
    which allows one to rewrite the pair $(\beta, \kappa)$ as
    \begin{equation}
        \beta = \delta \sqrt{2\epsilon - \epsilon^2},
        \quad
        \kappa = \sqrt{\frac{2 - \epsilon}{\epsilon}}.
    \end{equation}
    Thus, the system approaches degeneracy as \(\epsilon \to 0\), while simultaneously the condition number \(\kappa\) diverges to infinity.

    From our previous analysis, when the system tends to be degenerate, the non-normal gain tends to zero.
    However, here we face competing asymptotics: \(\beta \to 0\) and \(\kappa \to \infty\).
    It is therefore necessary to investigate which effect dominates.

    This \(\epsilon\)-parametrization was employed in \cite{Charan2020,Charan2021Remote} to analyze amplification as \(\epsilon \to 0^+\),
    considering sinusoidal forcing. In this setting, the degree of non-normality reads
    \begin{equation}
        K = \sqrt{\frac{\kappa^2 - 1}{2}} = \frac{1}{\sqrt{\epsilon}}.
    \end{equation}
    The non-normal gain is given by
    \begin{equation}
        G = K^2 \Delta v, \quad \text{with} \quad
        \Delta v = \frac{2\beta^2}{(\omega^2 - 1)^2 + \beta^2 + (\gamma \omega)^2}.
    \end{equation}
    Hence,
    \begin{equation}
        G(\epsilon) = \frac{2 \delta^2 (2 - \epsilon)}{(\omega^2 - 1)^2 + \delta^2 (2\epsilon - \epsilon^2) + (\gamma \omega)^2},
        \quad 1 \ge \epsilon \ge 0.
    \end{equation}

    This shows that the singularity in \(K\) as \(\epsilon \to 0\) is exactly canceled by the vanishing \(\beta^2\).  
    Consequently, the non-normal gain \(G(\epsilon)\) remains finite and smooth over the entire range \(\epsilon \in [0,1]\), with a minimum at \(\epsilon=0\) and a maximum at \(\epsilon=1\).

    Therefore, the \(\epsilon\)-parametrization, although useful for exploring criticality, does not allow one to observe high pseudo-criticality due to non-normality unless both the forcing frequency and damping satisfy the very restrictive conditions \(\omega=1\) and \(\gamma=0\) simultaneously.  
    In this exceptional case, \(G = K^2 = 1/\epsilon\) diverges, revealing a true non-normal amplification singularity as \(\epsilon \to 0^+\).

    Since these conditions are highly restrictive and unlikely to hold simultaneously in practice, it is nearly impossible to observe strong non-normal amplification solely through this parametrization.
    \newline

    A similar conclusion holds for Gaussian forcing, where
    \begin{equation}
        K = \frac{\kappa - \kappa^{-1}}{2} = \frac{1}{\sqrt{2\epsilon + \epsilon^2}},
    \end{equation}
    which also diverges like \(1/\sqrt{\epsilon}\) as \(\epsilon \to 0^+\).

    For both real (\(\chi=1\)) and complex conjugate (\(\chi = i\)) eigenvalues, the non-normal gain scales as
    \begin{equation}
        G \propto (K \beta)^2 = \delta^2 \frac{2 - \epsilon}{2 + \epsilon}, \quad 1 \ge \epsilon \ge 0,
    \end{equation}
    which remains finite for all \(\epsilon > 0\).

    As long as damping \(\gamma > 0\) is present, the singularity in \(K\) is effectively canceled in the gain \(G\). Thus, the system never achieves the full non-normal amplification unless \(\gamma \to 0\),
    a limit that corresponds physically to criticality rather than non-normality.
    \newline

    In conclusion, despite the parametrization \(\epsilon \sim 0\) producing a diverging condition number \(\kappa\) and degree of non-normality \(K\),
    the simultaneous vanishing of \(\beta\) means the system never becomes truly ``highly" non-normal.  
    Hence, the amplification mechanisms studied in \cite{Charan2020,Charan2021Remote} correspond primarily to classical critical behavior,
    not to non-normal amplification effects.

    \subsubsection{Conclusion}

    The reduced form proposed by \cite{Charan2020} to explain amplification mechanisms in frictional amorphous solids
    -- potentially shedding light on remote earthquake triggering \cite{Charan2021Remote} --
    indeed involves non-normality. However, the non-normal behavior is not fully exploited or manifested in terms of amplification.

    Up to a unitary transformation, the reduced form that we obtain exactly matches the \(\beta\)-form introduced in the present work,
    making it an ideal example to apply our formalism.

    The \(\epsilon\)-parametrization proposed by \cite{Charan2020,Charan2021Remote} describes the system approaching criticality,
    with scaling \(\beta \sim \sqrt{\epsilon}\) and \(K \sim 1/\sqrt{\epsilon}\).  
    At first glance, it appears that amplification will result from the divergence of the degree of non-normality \(K\) as \(\epsilon \to 0^+\).
    While it is true that $K$ diverges as \(\epsilon \to 0^+\), the actual non-normal amplification mechanism depends on the product \(K \beta\),
    which remains on the order of unity in this limit.
    Therefore, the \(\epsilon\)-parametrization obscures the presence of any strong non-normal amplification mechanism,
    effectively hiding an entire regime of dynamics where non-normal effects would dominate.

    This highlights that the usual emphasis on criticality and resonance in the study of complex systems does not capture the full picture of amplification mechanisms,
    and that considering non-normality explicitly is essential for a more complete understanding.

    \subsection{Four-Wave Mixing}

    Non-Hermitian physics studies systems governed by non-Hermitian Hamiltonians,
    which can have complex eigenvalues leading to phenomena such as exceptional points and non-conservative dynamics \cite{el2018non}.
    These systems exhibit unconventional transport properties,
    enabling novel mechanisms for enhancing engine performance through the interplay of gain,
    loss, and coherence \cite{ashida2020non}.

    Non-normality adds further complexity to non-Hermitian systems.
    Since non-normal operators have non-orthogonal eigenvectors, their eigenbasis is non-unitary,
    leading to transient dynamics characterized by temporary amplifications or decays beyond what eigenvalues alone predict \cite{Embree2005}.
    In strongly non-normal systems, such transient behavior is controlled by eigenvector alignments and operator structure, profoundly influencing short-term evolution \cite{troude2025}.

    We demonstrate that the non-normal amplification mechanisms described by the two \(\alpha/\beta\)-formalisms apply to the (Forward) Four-Wave Mixing (FFWM) experiment.
    FFWM is a nonlinear optical process in which four electromagnetic waves interact in a medium, generating new frequencies.
    In the setup of \cite{Yue2019}, FFWM occurs in a cold atomic ensemble with a double-\(\Lambda\) four-level configuration.
    It involves a pump laser (\(\omega_p\)), a coupling laser (\(\omega_c\)), and seeded Stokes (\(\omega_s\)) and anti-Stokes (\(\omega_{as}\)) fields.
    This system has been used to demonstrate Parity-Time (PT) symmetry breaking,
    which corresponds to a Hopf bifurcation.

    We show that the non-Hermitian Hamiltonian describing this FFWM setup is not only non-Hermitian but also non-normal,
    with the PT symmetry breaking fitting naturally into the non-normal \(\beta\)-framework.
    However, similar to frictional amorphous solids, non-normality is suppressed near the coalescence of eigenvalues.
    Further from the PT breaking point, the system can be described by the \(\alpha\)-formalism,
    where the system becomes highly non-normal as both the small linear Raman amplification \(l\) from the pump laser and the linear absorption \(a\) due to imperfect Electromagnetically Induced Transparency (EIT) vanish,
    allowing non-normal amplification to dominate.

    \subsubsection{System Dynamics}

    The FFWM setup can be modeled as an overdamped system described by the equation
    \begin{equation}    \label{eq:apx_ffwm}
        \frac{\partial}{\partial z}\x = \J \x + \g,
        \quad \text{where} \quad
        \J =
        \begin{pmatrix}
            -a + i \frac{\Delta k}{2} & i c \\
            -i c & l - i \frac{\Delta k}{2}
        \end{pmatrix},
    \end{equation}
    where \(\x\) is the state vector, and \(\g\) represents a source term,
    and $z$ is the axis along which the wave propagates.

    Here, 
    \begin{itemize}
        \item \(l\) denotes the small linear Raman amplification induced by the pump laser,
        \item \(a\) accounts for unavoidable linear absorption due to imperfect Electromagnetically Induced Transparency (EIT),
        stemming from nonzero dephasing rates between atomic states,
        \item \(\Delta k\) is the real phase mismatch between the interacting waves,
        \item and \(c\) characterizes the coupling strength in the system.
    \end{itemize}

    For simplicity, we consider all parameters \((a, l, c, \Delta k)\) to be real-valued, ensuring system stability.
    Further details on these parameters and their physical origins can be found in \cite{Yue2019}.

    \subsubsection{Non-Normal Decomposition}

    To characterize the degree of non-normality \(K\) and the spectral properties of the system matrix \(\J\) in \eqref{eq:apx_ffwm},
    we start by computing its eigenvalues and eigenvectors.

    The eigenvalues \(\lambda_\pm\) are given by
    \begin{equation}
        \lambda_\pm = -\frac{a - l}{2} \pm \frac{i}{2} \sqrt{(\Delta k + i(a + l))^2 - 4 c^2},
    \end{equation}
    and the corresponding normalized eigenvectors \(\pp_\pm\) are
    \begin{equation}
        \pp_\pm = \frac{1}{\sqrt{1 + |\nu_\pm|^2}} 
        \begin{pmatrix} 1 \\ \nu_\pm \end{pmatrix},
        \quad \text{where} \quad
        \nu_\pm = \nu \pm \sqrt{\nu^2 - 1}, \quad \nu = \frac{\Delta k + i (a + l)}{2 c}.
    \end{equation}
    The parameter \(\nu\) controls the non-normality of the system,
    as the eigenbasis transformation matrix is \(\PP = (\pp_+, \pp_-)\).

    The condition number \(\kappa\) of \(\PP\), which quantifies non-normality, can be expressed as
    \begin{equation}
        \kappa = \sqrt{\frac{1 + |\pp_+^\dag \pp_-|}{1 - |\pp_+^\dag \pp_-|}},
    \end{equation}
    where the overlap between eigenvectors is
    \begin{equation}
        |\pp_+^\dag \pp_-| = \frac{|1 + \nu_+^* \nu_-|}{\sqrt{\left(1 + |\nu_+|^2\right)\left(1 + |\nu_-|^2\right)}}.
    \end{equation}
    To simplify the denominator, consider
    \begin{equation}
        \left(1 + |\nu_+|^2\right)\left(1 + |\nu_-|^2\right) = 1 + |\nu_+ \nu_-|^2 + |\nu_+|^2 + |\nu_-|^2,
    \end{equation}
    and use the identities
    \begin{equation}
        |\nu_\pm|^2 = |\nu|^2 + |\sqrt{\nu^2 - 1}|^2 \pm 2 \mathrm{Re}\left(\nu^* \sqrt{\nu^2 - 1}\right),
    \end{equation}
    along with
    \begin{align}
        |1 + \nu_+^* \nu_-|^2
        &= 1 + |\nu_+ \nu_-|^2 + 2 \mathrm{Re}(\nu_+^* \nu_-) \\
        &= 1 + |\nu_+ \nu_-|^2 + 2 \left(|\nu|^2 - |\sqrt{\nu^2 - 1}|^2\right).
    \end{align}
    Rearranging yields
    \begin{equation}
        1 + |\nu_+ \nu_-|^2 = |1 + \nu_+^* \nu_-|^2 + 2\left(|\sqrt{\nu^2 - 1}|^2 - |\nu|^2\right),
    \end{equation}
    so that
    \begin{equation}
        \left(1 + |\nu_+|^2\right)\left(1 + |\nu_-|^2\right) = |1 + \nu_+^* \nu_-|^2 + 4 |\nu^2 - 1|.
    \end{equation}
    Therefore, the eigenvector overlap can be compactly written as
    \begin{equation}
        |\pp_+^\dag \pp_-| = \frac{|1 + \nu_+^* \nu_-|}{\sqrt{|1 + \nu_+^* \nu_-|^2 + 4 |\nu^2 - 1|}},
    \end{equation}
    and the condition number becomes
    \begin{equation}
        \kappa = \sqrt{
            \frac{
                \sqrt{|1 + \nu_+^* \nu_-|^2 + 4 |\nu^2 - 1|} + |1 + \nu_+^* \nu_-|
            }{
                \sqrt{|1 + \nu_+^* \nu_-|^2 + 4 |\nu^2 - 1|} - |1 + \nu_+^* \nu_-|
            }
        }.
    \end{equation}
    From this expression, it follows that \(\kappa\) diverges as \(\nu \to \pm 1\),
    indicating a highly non-normal regime near these parameter values.

    \subsubsection{PT Symmetry Breaking and Non-Normality}

    Recall the eigenvalues and condition number of the system
    \begin{equation}
        \lambda_\pm = -\frac{a - l}{2} \pm i c \sqrt{\nu^2 - 1}, \quad
        \kappa = \sqrt{\frac{\sqrt{|1+\nu_+^* \nu_-|^2 + 4|\nu^2 - 1|} + |1 + \nu_+^* \nu_-|}{\sqrt{|1+\nu_+^* \nu_-|^2 + 4|\nu^2 - 1|} - |1 + \nu_+^* \nu_-|}},
    \end{equation}
    where
    \begin{equation}
        \nu_\pm = \nu \pm \sqrt{\nu^2 - 1}, \quad \nu = \frac{\Delta k + i(a + l)}{2 c}.
    \end{equation}
    The system becomes highly non-normal as \(\nu \to \pm 1\), which also corresponds to a bifurcation point known as PT symmetry breaking in this class of systems.
    At these points, eigenvalues coalesce, making the system degenerate and causing the divergence in the condition number to be canceled, as observed in the frictional amorphous solids case.

    Thus, PT symmetry breaking fits naturally into the \(\beta\)-framework:
    it occurs when eigenvalues become degenerate,
    explaining why non-normal amplification tends to be suppressed near the PT symmetry breaking point.
    \newline

    To explore parameter regimes that allow non-normal amplification away from PT symmetry breaking,
    consider \(\Delta k, c > 0\), and focus on approaching \(\nu \to 1\). Define the small deviation
    \begin{equation}
        \delta \nu = \nu - 1 = \frac{\Delta k - 2 c}{2 c} \left[1 + i \frac{a + l}{\Delta k - 2 c}\right].
    \end{equation}
    For the condition number to diverge, \(\Delta k \approx 2 c\) is required. To remain sufficiently away from the PT-breaking point, we impose
    \begin{equation}
        |\Delta k - 2 c| \gg a + l.
    \end{equation}
    Under these conditions, the eigenvalues simplify to
    \begin{equation}
        \lambda_\pm = \sqrt{\left(\frac{\Delta k}{2}\right)^2 - c^2} \left[-\alpha \pm i + \mathcal{O}\left(\frac{a + l}{\Delta k - 2 c}\right)\right],
        \quad \alpha = \frac{a - l}{\sqrt{\Delta k^2 - 4 c^2}}.
    \end{equation}
    Rescaling the length as
    \(
        z \to z\sqrt{\left(\frac{\Delta k}{2}\right)^2 - c^2},
    \)
    we recover the \(\alpha\)-framework. The condition number then behaves as
    \begin{equation}
        \kappa = \sqrt{\left|\frac{\Delta k + 2 c}{\Delta k - 2 c}\right|} + \mathcal{O}\left(\left(\frac{a + l}{\Delta k - 2 c}\right)^2\right).
    \end{equation}
    Assuming Gaussian forcing, the relevant degree of non-normality is
    \begin{equation}
        K = \frac{1}{2} |\kappa - \kappa^{-1}| = \frac{\max\{\Delta k, 2 c\}}{\sqrt{\Delta k^2 - 4 c^2}} + \mathcal{O}\left(\left(\frac{a + l}{\Delta k - 2 c}\right)^2\right).
    \end{equation}
    The critical non-normality threshold for complex conjugate eigenvalues is given by
    \begin{equation}
        K_c = \sqrt{\frac{\alpha}{\sqrt{\alpha^2 + 1} - \alpha}} = \sqrt{\alpha} + \mathcal{O}(\alpha^{3/2}),
    \end{equation}
    so to observe significant non-normal behavior, one requires
    \(
        K > K_c,
    \)
    and for large transient amplification,
    \(
        K \gg K_c,
    \)
    which translates to the condition
    \begin{equation}
        \max\{\Delta k, 2 c\} \gg \sqrt{a - l} \, (\Delta k^2 - 4 c^2)^{1/4}.
    \end{equation}
    In summary, the hierarchy of hyper-parameters required for significant non-normal amplification in the FFWM experiment is
    \begin{equation}
        \Delta k \approx 2 c \gg |\Delta k - 2 c| \gg a + l > a - l > 0, \quad a > l > 0.
    \end{equation}

    \subsubsection{Conclusion}

    The general theoretical framework developed in this paper can be successfully applied to non-Hermitian quantum systems.  
    Using the specific example of the Forward Four-Wave Mixing (FFWM) experiment,  
    we retained the effects of damping arising from the small linear Raman amplification and the linear absorption due to Electromagnetically Induced Transparency (EIT).

    We demonstrated that the PT symmetry breaking phenomenon fits naturally within the non-normal \(\beta\)-framework,  
    as it corresponds to approaching system degeneracy, which effectively suppresses non-normal amplification.

    To recover significant non-normal amplification, we proposed exploring the parameter space beyond the PT symmetry breaking point,  
    transforming the description from the \(\beta\)-framework to the \(\alpha\)-framework.  
    This allows us to establish a hierarchy among the control parameters that governs the emergence of strong non-normal behavior in this system.

    This approach opens new avenues for understanding and controlling amplification mechanisms in non-Hermitian quantum systems.

    \subsection{Conclusion}

    We have applied the non-normal unifying framework for amplification mechanisms to existing models from classical solid dynamics and quantum optics.  
    Our analysis shows that models focused solely on bifurcations may overlook important aspects of non-normality.  
    By exploring the parameter space more broadly, we identified where significant non-normal amplification can occur.

    This approach enables researchers to directly assess whether non-normality could underlie phenomena such as earthquake triggering or noise amplification in quantum optical systems.  
    Thus, it provides a valuable tool to deepen our understanding of amplification mechanisms across natural sciences.

\end{document}